\documentclass[twocolumn,twoside,slac_two]{revtex4}
\usepackage{graphicx}
\usepackage{fancyhdr}
\newcommand{\GeV}{\rm{GeV}}

\usepackage{units}
\RequirePackage{xspace}

\usepackage{relsize}
\def\babar{\mbox{\slshape B\kern-0.1em{\smaller A}\kern-0.1em
    B\kern-0.1em{\smaller A\kern-0.2em R}}\xspace}

\def\epem       {\ensuremath{e^+e^-}\xspace}

\def\ccbar {\ensuremath{c\overline c}\xspace}

\def\piz   {\ensuremath{\pi^0}\xspace}

\def\pip   {\ensuremath{\pi^+}\xspace}

\def\Kbar  {\kern 0.2em\overline{\kern -0.2em K}{}\xspace}

\def\Kz    {\ensuremath{K^0}\xspace}
\def\Kzb   {\ensuremath{\Kbar^0}\xspace}
\def\KzKzb {\ensuremath{\Kz \kern -0.16em \Kzb}\xspace}
\def\Kp    {\ensuremath{K^+}\xspace}
\def\Km    {\ensuremath{K^-}\xspace}

\def\KpKm  {\ensuremath{\Kp \kern -0.16em \Km}\xspace}

\def\Dbar    {\kern 0.2em\overline{\kern -0.2em D}{}\xspace}

\def\Dz      {\ensuremath{D^0}\xspace}
\def\Dzb     {\ensuremath{\Dbar^0}\xspace}
\def\DzDzb   {\ensuremath{\Dz {\kern -0.16em \Dzb}}\xspace}
\def\Dp      {\ensuremath{D^+}\xspace}
\def\Dm      {\ensuremath{D^-}\xspace}

\def\DpDm    {\ensuremath{\Dp {\kern -0.16em \Dm}}\xspace}

\def\Dstarp  {\ensuremath{D^{*+}}\xspace}

\def\Ds      {\ensuremath{D^+_s}\xspace}

\def\Dss     {\ensuremath{D^{*+}_s}\xspace}

\def\B       {\ensuremath{B}\xspace}
\def\Bbar    {\kern 0.18em\overline{\kern -0.18em B}{}\xspace}

\def\BB      {\ensuremath{B\Bbar}\xspace} 
\def\Bz      {\ensuremath{B^0}\xspace}
\def\Bzb     {\ensuremath{\Bbar^0}\xspace}
\def\BzBzb   {\ensuremath{\Bz {\kern -0.16em \Bzb}}\xspace}
\def\Bu      {\ensuremath{B^+}\xspace}
\def\Bub     {\ensuremath{B^-}\xspace}

\def\BpBm    {\ensuremath{\Bu {\kern -0.16em \Bub}}\xspace}

\def\BorBbar    {\kern 0.18em\optbar{\kern -0.18em B}{}\xspace}
\def\DorDbar    {\kern 0.18em\optbar{\kern -0.18em D}{}\xspace}
\def\KorKbar    {\kern 0.18em\optbar{\kern -0.18em K}{}\xspace}

\mathchardef\Upsilon="7107
\def\Y#1S{\ensuremath{\Upsilon{(#1S)}}\xspace}

\def\FourS {\Y4S}

\mathchardef\Deltares="7101
\mathchardef\Xi="7104
\mathchardef\Lambda="7103
\mathchardef\Sigma="7106
\mathchardef\Omega="710A

\def\Deltabar{\kern 0.25em\overline{\kern -0.25em \Deltares}{}\xspace}
\def\Lbar{\kern 0.2em\overline{\kern -0.2em\Lambda\kern 0.05em}\kern-0.05em{}\xspace}
\def\Sigbar{\kern 0.2em\overline{\kern -0.2em \Sigma}{}\xspace}
\def\Xibar{\kern 0.2em\overline{\kern -0.2em \Xi}{}\xspace}
\def\Obar{\kern 0.2em\overline{\kern -0.2em \Omega}{}\xspace}
\def\Nbar{\kern 0.2em\overline{\kern -0.2em N}{}\xspace}
\def\Xb{\kern 0.2em\overline{\kern -0.2em X}{}\xspace}

\def\BR         {{\ensuremath{\cal B}\xspace}}

\newcommand{\tev}{\ensuremath{\mathrm{\,Te\kern -0.1em V}}\xspace}
\newcommand{\gev}{\ensuremath{\mathrm{\,Ge\kern -0.1em V}}\xspace}
\newcommand{\mev}{\ensuremath{\mathrm{\,Me\kern -0.1em V}}\xspace}
\newcommand{\kev}{\ensuremath{\mathrm{\,ke\kern -0.1em V}}\xspace}
\newcommand{\ev}{\ensuremath{\mathrm{\,e\kern -0.1em V}}\xspace}
\newcommand{\gevc}{\ensuremath{{\mathrm{\,Ge\kern -0.1em V\!/}c}}\xspace}
\newcommand{\mevc}{\ensuremath{{\mathrm{\,Me\kern -0.1em V\!/}c}}\xspace}
\newcommand{\gevcc}{\ensuremath{{\mathrm{\,Ge\kern -0.1em V\!/}c^2}}\xspace}
\newcommand{\mevcc}{\ensuremath{{\mathrm{\,Me\kern -0.1em V\!/}c^2}}\xspace}

\def\fb   {\ensuremath{\mbox{\,fb}}\xspace}

\def\mus  {\ensuremath{\rm \,\mus}\xspace}

\def\mus        {\ensuremath{\,\mu{\rm s}}\xspace}    

\def\to                 {\ensuremath{\rightarrow}\xspace}

\def\pep2{PEP-II}

\def\gsim{{~\raise.15em\hbox{$>$}\kern-.85em
          \lower.35em\hbox{$\sim$}~}\xspace}
\def\lsim{{~\raise.15em\hbox{$<$}\kern-.85em
          \lower.35em\hbox{$\sim$}~}\xspace}

\def\qsq                {\ensuremath{q^2}\xspace}

\xspace
\newcommand{\fds}{\ensuremath{f_{D_s}}\xspace}

\def\jetset74   {\mbox{\tt Jetset \hspace{-0.5em}7.\hspace{-0.2em}4}\xspace}

\def\SigFitYield{489}
\def\SigFitYieldError{55}

\def\PWRValStat{0.018}
\def\PWRVal{0.143}

\def\FDsbabarValNorm{14}
\def\FDsbabarValSyst{7}
\def\FDsbabarValStat{17}
\def\FDsbabarVal{283}

\def\BRbabarValNorm{0.66}
\def\BRbabarValSyst{0.26}
\def\BRbabarValStat{0.83}
\def\BRbabarVal{6.74}

\def\SigFitYieldGOF{8.9}

\def\dsPhiPiDataFitYield{2093}
\def\dsPhiPiDataFitYieldError{99}
\def\dsPhiPiDataGOF{25.0}

\def\dstomunu{\ensuremath{\Ds\to\mu^+\nu_\mu}\xspace}
\def\dstophipi{\ensuremath{\Ds\to\phi\pi^+}\xspace}
\def\dsstodstophipi{\ensuremath{\Dss\to\gamma\Ds\to\gamma\phi\pi^+}\xspace}

\newcommand{\thl}{\theta_e}
\newcommand{\GeVcd}{\rm{GeV/}c^2}
\newcommand{\beq}{\begin{eqnarray}}
\newcommand{\eeq}{\end{eqnarray}}
\newcommand{\Do}{D^0}
\newcommand{\GeVc}{\rm{GeV/}c}
\newcommand{\pipl}{\pi^{+}}

\begin{document}

\title{Leptonic and semileptonic $D$ and $D_s$ decays at B-factories}

%

\author{L. Widhalm (Belle collaboration)}
\affiliation{Institute of High-Energy Physics, Austrian Academy of Sciences, Vienna 1050, Austria}

\begin{abstract}
Recent measurements of branching fractions, form factors and decay constants of leptonic and semileptonic decays of $D_{(s)}$-mesons acquired at experiments running at the $\Upsilon(4S)$ resonance energy are reviewed.
\end{abstract}

\maketitle

\thispagestyle{fancy}


\section{Introduction}

One of the important goals of particle physics is the precise measurement and understanding of the Cabibbo-Kobayashi-Maskawa (CKM) Matrix. To interpret results from B-factory experiments such as \babar~\cite{ref:babar} and Belle~\cite{Belle}, theoretical calculations of form factors and decay constants (usually based on lattice gauge theory, see e.g.~\cite{Kronfeld:2006sk}) are needed. It is necessary to have accurate measurements in the charm sector to check (and allow further tuning of) theoretical methods and predictions.


Due to their relative abundance and simplified theoretical treatment, (semi)leptonic decays of $D$ or $D_s$ mesons are a favored means of
determining the weak interaction couplings of quarks within the standard model.

\vspace{3mm}
This review concentrates on experimental results for such decays achieved at experiments running at the $\Upsilon(4S)$ resonance threshold, namely the \babar and Belle experiments. It uses adapted excerpts from the cited Belle and \babar publications.

\section{The Experiments}


The \babar detector~\cite{ref:babar} reconstructs
charged particles by matching hits in 
the 5-layer double-sided silicon vertex tracker (SVT) 
with track elements in the 40-layer drift chamber (DCH), which is
filled with a gas mixture of helium and isobutane.
Slow particles which do not leave enough hits  
in the DCH due to the bending in the $1.5$-T
magnetic field, are reconstructed in the SVT.
Charged hadron identification is performed combining the measurements of 
the energy deposition in the SVT and in the DCH with the information from the
Cherenkov detector (DIRC). Photons are detected and measured in the 
CsI(Tl) electro-magnetic calorimeter (EMC). 
Electrons are identified by the ratio of the track momentum to the
associated energy deposited in the EMC, the transverse profile of the shower,
the energy loss in the DCH, and the Cherenkov angle in the DIRC.
Muons are identified in the instrumented flux return, composed
of resistive plate chambers interleaved with layers of steel and brass.


The Belle detector~\cite{Belle} is a large-solid-angle magnetic spectrometer that
consists of a silicon vertex detector (SVD), a 50-layer central drift chamber (CDC), an array of aerogel threshold Cherenkov counters (ACC), a barrel-like arrangement of time-of-flight scintillation counters (TOF), and an electromagnetic calorimeter comprised of CsI(Tl) crystals (ECL) located inside a superconducting solenoid coil that provides a 1.5~T magnetic field.  An iron flux-return located outside of the coil is instrumented to detect $K_L^0$ mesons and to identify muons (KLM).  The detector is described in detail elsewhere~\cite{Belle}.  Two inner detector configurations were used. A 2.0 cm beampipe and a 3-layer silicon vertex detector were used for the first sample of $156$ fb$^{-1}$, while a 1.5 cm beampipe, a 4-layer silicon detector and a small-cell inner drift chamber were used to record the remaining $392$ fb$^{-1}$~\cite{svd2}.  

\section{Semileptonic Decays to Scalar Mesons}

Form factors from $D$ meson semileptonic decay have been calculated using
lattice QCD techniques~\cite{ref:unquenched,ref:unquenched2,ref:quenched}.
In the theoretical description, the differential decay width is dominated by 
the form factor $f_+(q^2)$~\cite{ref:2a}, where $q^2$ is the invariant mass of the lepton pair.
Up to order $m_{\mathrm \ell}^2$ it is given by
\begin{eqnarray}
\frac{d\Gamma^{K(\pi)}}{dq^2} & = & \frac{G_F^2|V_{cs(d)}|^2}{24\pi^3}{|f_+^{K(\pi)}(q^2)|^2}p_{\mathrm K(\pi)}^3
\label{equ:ff}
\end{eqnarray}
where $p_{\mathrm K(\pi)}$ is the magnitude  of the meson 3-momentum in the $\bar D_\mathrm{sig}^0$ rest frame.

In the {\it modified pole model}~\cite{ref:bk}, the form factor $f_+$ is described as
\begin{eqnarray}
  f_+(q^2) &= & \frac{f_+(0)}{(1-q^2/m_{\mathrm{pole}}^2)(1-\alpha_p q^2/m_{\mathrm{pole}}^2)},
\end{eqnarray}
with the pole masses predicted as
$m(D_s^*) = 2.11$~GeV/$c^2$ (for $\bar D_\mathrm{sig}^0 \rightarrow K^+ \ell^- \nu$) and
$m(D^*) = 2.01$~GeV/$c^2$ (for $\bar D_\mathrm{sig}^0 \rightarrow \pi^+ \ell^- \nu$). 
Setting $\alpha_p = 0$ leads to the {\it simple pole model}~\cite{ref:2a}.

A model independent description of the form factor has been studied in \cite{ref:hill1}. The most general expressions of the form factor $f_+(q^2)$ are analytic
functions satisfying the dispersion relation:
\begin{equation}
f_+(q^2) = \frac{Res (f_+)_{q^2=m^2_{D_s^*}}}{ m_{D_s^*}^2-q^2}
+ \frac{1}{\pi} \int_{t_+}^{\infty} dt \frac{\Im{f_+(t)}}{t-q^2-i\epsilon}.
\end{equation}
The only singularities in the  complex $t\equiv q^2$ plane
originate from the interaction of the charm and the strange quarks
in vector states. They are a pole, situated at the $D_s^*$ mass squared
and a cut, along the positive real axis, starting at threshold ($t_+=(m_D+m_K)^2$)
for $D^0K^-$ production.

This cut $t$-plane can be mapped onto the open unit disk with center at $t=t_0$
using the variable:
\begin{equation} z(t,t_0) = \frac{\sqrt{t_+-t}-\sqrt{t_+-t_0}}{\sqrt{t_+-t}+\sqrt{t_+-t_0}}.
\end{equation}
In this variable, the physical region for the semileptonic decay 
corresponds to the small (real) range between $\pm z_{\rm max}=\pm 0.051$. 
The $z$ expansion of $f_+$,
\begin{equation}
f_+(t) \propto \sum_{k=0}^{\infty}a_k(t_0)~ z^k(t,t_0),
\end{equation}
is thus expected to converge quickly. 

\subsection{$D \to (K/\pi) (e/\mu) \nu_{(e/\mu)}$ at Belle}

Belle has measured the
absolute branching fractions and form factors of $D^0 \to K^- l^+ \nu_l$ and $D^0 \to \pi^- l^+
\nu_l$ ($l=e,\mu$)~\cite{ref:belle},
using a novel reconstruction
method with better $q^2$ resolution than in previous experiments.
The analysis is based on
data corresponding to a total integrated luminosity of 282 
fb$^{-1}$.

To achieve good resolution in the neutrino momentum and $q^2$, the $D^0$ are tagged by fully reconstructing the remainder of the event.
The studied events are of the type $e^+e^-\rightarrow D_\mathrm{tag}^{(*)}D_\mathrm{sig}^{*-}X$ $\{D_\mathrm{sig}^{*-}\rightarrow \bar D_\mathrm{sig}^0\pi_s^-\}$, where $X$ may include additional $\pi^\pm$, $\pi^0$, or $K^\pm$ mesons (inclusion of charge-conjugate states is implied throughout this paper).
The $D_\mathrm{tag}^{(*)}$ is reconstructed in the modes $D^{*+}\rightarrow D^0\pi^+, D^+\pi^0$ and $D^{*0}\rightarrow D^0\pi^0, D^0\gamma$, with $D^{+/0}\rightarrow K^-(\mathrm{n}\pi)^{++/+}$ \{$\mathrm{n}=1,2,3$\}.  
Each $D_\mathrm{tag}$ and $D_\mathrm{tag}^*$ candidate is subjected to a mass-constrained vertex fit to improve the momentum resolution.
The 4-momentum of $D_\mathrm{sig}^{*-}$ is found by energy-momentum conservation, assuming a $D_\mathrm{tag}^{(*)}D_\mathrm{sig}^{*-}X$ event.
Its resolution is improved by subjecting it to a fit of the $X$ tracks and the $D_\mathrm{tag}^{(*)}$ momentum, constrained to originate at the run-by-run average collision point, while the invariant mass is constrained to the nominal mass of a $D^{*-}$.
Candidates for $\pi_s^-$ are selected from among the remaining tracks, and for each the candidate $\bar D_\mathrm{sig}^0$ 4-momentum is calculated from that of the $D_\mathrm{sig}^{*-}$ and $\pi_s^-$. 
The momentum is then adjusted by a kinematic fit constraining the candidate mass to that of the $D^0$. For this fit, the decay vertex of the $\bar D_\mathrm{sig}^0$ has been estimated by extrapolating from the collision point in the direction of the $\bar D_\mathrm{sig}^0$ momentum assuming the average decay length.

Background lying under the $\bar D_\mathrm{sig}^0$ mass peak (i.e. fake-$\bar D_\mathrm{sig}^0$)
is estimated using a wrong sign (WS) sample
where the tag- and signal-side $D$ candidates have the same flavor
 ($\bar D_\mathrm{tag}$ instead of $D_\mathrm{tag}$).
A MC study (including $\Upsilon(4S) \rightarrow B\overline{B}$ and
continuum ($q\bar{q}$, where $q = c$, $s$, $u$, $d$) events~\cite{ref:bellegen,ref:bellemc}) has found that this sample can
properly model the shape of background except for a small contribution from real 
$\bar D_\mathrm{sig}^0$ decays ($\approx 2\%$) from interchange between particles used for the tag due to particle misidentification. 
Background from fake $\bar D_\mathrm{sig}^0$ is subtracted normalizing this shape in a sideband region $1.84-1.85$~GeV/$c^2$, yielding $56461\pm309_\mathrm{stat}\pm830_\mathrm{syst}$ signal $\bar D_\mathrm{sig}^0$ tags. 

Within this sample of $\bar D_\mathrm{sig}^0$ tags, the semileptonic decay $\bar D_\mathrm{sig}^0\rightarrow K^+(\pi^+)\ell^-\bar\nu$ is reconstructed with $K^+(\pi^+)$ and $\ell^-$ candidates from among the remaining tracks.
The neutrino candidate 4-momentum is reconstructed by energy-momentum conservation, and its invariant mass squared, $m_{\nu}^2$, is required to satisfy $|m_{\nu}^2|<0.05~{\rm GeV}^2/c^4$.

Multiple candidates still remain in one third of $\bar D_\mathrm{sig}^0$ tags, and in
about one quarter of the semileptonic sample. In these cases all
candidates are saved and given equal weights such that each event has a total weight of $1$.

\begin{table*}[t]
\caption{Belle: Yields in data, estimated backgrounds, extracted signal yields and branching fractions, where for the latter two, the first uncertainty is statistical and second is systematic; small differences in the numbers are due to rounding.}
{\begin{tabular}{@{}lrrrrr@{}} \toprule
channel &
\multicolumn{1}{r}{full $\overline{D}^0_\mathrm{sig}$} &
\multicolumn{1}{r}{$K^+e^-\nu_e$} &
\multicolumn{1}{r}{$K^+\mu^-\nu_\mu$} &
\multicolumn{1}{r}{$\pi^+e^-\nu_e$} &
\multicolumn{1}{r}{$\pi^+\mu^-\nu_\mu$} \\
\hline
yield & $95250$ & $1349$ & $1333$ & $152$ & $141$ \\
fake $\overline{D}^0_\mathrm{sig}$ &
$38789$ & $12.6$ &  $12.2$ &  $12.3$ &  $12.5$ \\  
semileptonic & n/a & $6.7$ & $10.0$ & $11.7$ & $12.6$ \\
hadronic & n/a & $11.9$ & $62.1$ & $1.8$ & $9.7$ \\
\hline
signal & $56461$ & $1318$ & $1249$ & $126$ & $106$ \\
stat. error & 309 & 37 & 37 & 12 & 12 \\
syst. error & 830 & 7 & 25 & 3 & 6 \\
\multicolumn{2}{r}{} &
\multicolumn{1}{r}{} &
\multicolumn{1}{r}{} &
\multicolumn{1}{r}{} &
\multicolumn{1}{r}{} \\
\multicolumn{2}{r}{Branching Fraction ($10^{-4}$)} &
\multicolumn{1}{r}{$345 \pm 10 \pm 19$} &
\multicolumn{1}{r}{$345 \pm 10 \pm 21$} &
\multicolumn{1}{r}{$27.9 \pm 2.7 \pm 1.6$} &
\multicolumn{1}{r}{$23.1 \pm 2.6 \pm 1.9$} \\
\multicolumn{2}{r}{($e$ and $\mu$ channels, average)} &
\multicolumn{2}{c}{$345 \pm 7 \pm 20$} &
\multicolumn{2}{c}{$25.5 \pm 1.9 \pm 1.6$} \\
\end{tabular}
\label{tab:events}}
\end{table*}

\begin{figure}[t]
\begin{center}
\includegraphics[width=80mm]{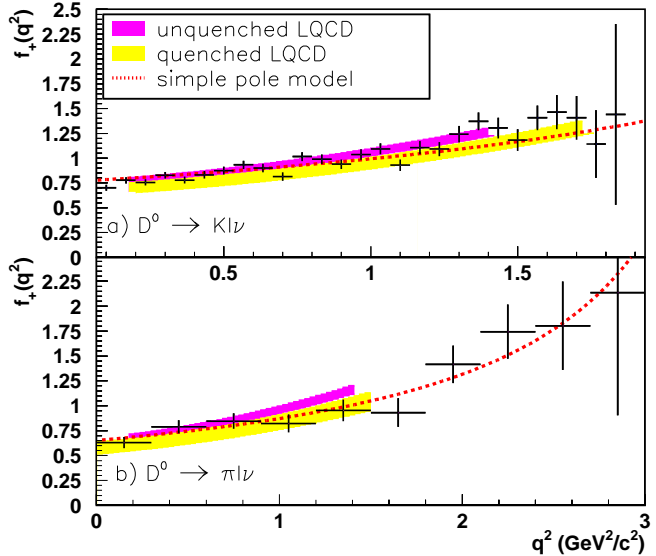}
\end{center}
\caption{Belle: Form factors for (a) $D^0 \rightarrow K^- \ell^+ \nu$, in $q^2$ bins of $0.067\ \mbox{GeV}^2/c^2$ and (b) $D^0 \rightarrow \pi^- \ell^+ \nu$, in $q^2$ bins of $0.3\ \mbox{GeV}^2/c^2$.
Overlaid are the
predictions of the simple pole model using the physical pole mass
(dashed), and a quenched (yellow) and unquenched (purple)
LQCD calculation.  Each LQCD curve is  obtained by fitting a parabola
to values calculated at specific $q^2$ points. The
shaded band reflects the theoretical uncertainty and is shown within the range of $q^2$ for which calculations are reported. 
\label{fig:X}}
\end{figure}

The contribution from fake $\bar D_\mathrm{sig}^0$ in the sample of semileptonic decay candidates is estimated using the $\bar D_\mathrm{sig}^0$ invariant mass WS shape of the $\bar D_\mathrm{sig}^0$ tag sample, normalized in the previously defined sideband region. 
Backgrounds from semileptonic decays with either an
incorrectly identified meson or where additional mesons are lost in
reconstruction are highly suppressed by the good
neutrino mass resolution. For $\bar D_\mathrm{sig}^0 \rightarrow \pi^+ \ell^- \nu$ the most significant
background is $\bar D^0 \rightarrow K^+ \ell^- \nu$
amounting to $6\%-8\%$ of the total yield. It
was estimated using the reconstructed $\bar D^0 \to K^+\ell^-\nu$ decays in data,
reweighted with the (independently measured) probability of kaons to fake pions. Smaller backgrounds from $\bar D^0 \rightarrow K^{*+} \ell^- \nu$ and
$\bar D^0 \rightarrow \rho^+ \ell^- \nu$ decays amounting to $0.8\%-0.9\%$ were measured by normalizing MC to
data in the upper sideband region $m^2_\nu>0.3\ \mbox{GeV}^2/c^4$, which is dominated by these channels.  For $\bar D_\mathrm{sig}^0 \rightarrow K^+ \ell^- \nu$, decays
of $\bar D^0 \rightarrow K^{*+} \ell^- \nu$ contribute at the level of $0.5\%-0.8\%$, measured using a sideband evaluation as described above, while background from
$\bar D^0 \to \pi^+ \ell^- \nu$ and $\bar D^0 \to \rho^+ \ell^- \nu$ was found to be negligible ($< 0.07\%$ of the total yield).
Background from $\bar D_\mathrm{sig}^0$ decays to hadrons, where a
hadron is mis-identified as a lepton,
is measured with an opposite sign (OS) sample, where the lepton charge is opposite to that of the $D^{*-}_\mathrm{sig}$ slow pion. 
Note that the
signal is extracted from the same sign (SS) sample.
In contrast to the SS sample, the OS sample has no
signal or semileptonic backgrounds; fake $\bar D^0_\mathrm{sig}$ are subtracted in the same manner described previously.
Assigning well identified pion and kaon tracks a lepton mass, pure background $m_\nu^2$ distributions are constructed in both SS and OS, which are labelled $f_m^\mathrm{SS}$ and $f_m^\mathrm{OS}$, $m=K,\pi$.
A fit of the weights $a_K$ and $a_\pi$ of the components $f_K^\mathrm{OS}$
and $f_\pi^\mathrm{OS}$ in the $m_\nu^2$ distribution of the OS data
sample is performed, and the hadronic background in the SS data
sample is calculated as $(a_K f_K^\mathrm{SS} + a_\pi f_\pi^\mathrm{SS})$, utilizing the fact that the hadron misidentification rate does not depend on the charge correlation defining SS and OS. The method has been
validated using MC samples. As the muon fake rate is about an order of
magnitude larger than that for electrons, this background is much more
significant for muon modes. 
The signal yields and estimated backgrounds are summarized in the upper part of Table~\ref{tab:events}. 

Efficiencies depend strongly on $n_X$, defined as the number of $\pi^{\pm(0)}$ and $K^\pm$ mesons assigned to $X$ (in $e^+e^-\rightarrow D_\mathrm{tag}^{(*)}D_\mathrm{sig}^{*-}X$), and are determined with MC; 
differences in the $n_X$ distribution between MC and data give rise 
to a further $(+1.9\pm3.9)\%$ correction.
Applying these corrections, the absolute branching fractions (normalized to the total number of $\bar D_\mathrm{sig}^0$ tags) summarized in the lower part of Table \ref{tab:events} are obtained.

The resolution in $q^2$ of semileptonic decays is found to be $\sigma_{q^2}=0.0145\pm 0.0007_\mathrm{stat}$~GeV$^2/c^2$ in MC signal events.
This is much smaller than statistically reasonable bin widths, which have been chosen as $0.067\ (0.3)\ \mbox{GeV}^2/c^2$ for kaon (pion) modes, and hence 
no unfolding is necessary.
Bias in the measurement of $q^2$ that may arise due to events where 
the lepton and meson are interchanged, a double mis-assignment, was 
checked with candidate $\bar D_\mathrm{sig}^0 \to K^+\ell^-\nu$ events and found to be negligible.  
The differential decay width is bin-by-bin background subtracted and efficiency
corrected, using the same methods described previously. 

The measured $q^2$ distribution
is fitted with $2$ free parameters to the predicted 
differential decay width $d\Gamma/dq^2$ of the pole models
with $f_+(0)$ being one of the parameters, and either $m_{\mathrm{pole}}$ (setting $\alpha_p=0$) or $\alpha_p$ (assuming the theoretical pole) the other. Binning effects are accounted for by averaging the
model functions within individual $q^2$ bins.
The fit to the simple pole model yields
$m_\mathrm{pole}(K^- \ell^+ \nu) = 1.82\pm0.04_\mathrm{stat}\pm0.03_\mathrm{syst}$~GeV/$c^2$ ($\chi^2/\mbox{ndf}=34/28$)
and $m_\mathrm{pole}(\pi^- \ell^+ \nu) = 1.97\pm0.08_\mathrm{stat}\pm0.04_\mathrm{syst}$~GeV/$c^2$ ($\chi^2/\mbox{ndf}=6.2/10$).
While the pole mass for the $\pi \ell \nu$ decay agrees within errors with the predicted value, $m(D^*)$, the more accurate fit of $m_{\mathrm{pole}}(K \ell \nu)$ is 
several standard deviations below $m(D_s^*)$.
In the modified pole model, $\alpha_p$ describes this deviation of the real poles from the $m(D_\mathrm{(s)}^*)$ masses. Fixing these masses to their known experimental values, a fit of  $\alpha_p$ yields 
$\alpha_p (D^0 \to K^- \ell^+ \nu) = 0.52 \pm 0.08_\mathrm{stat} \pm 0.06_\mathrm{syst}$ ($\chi^2/\mbox{ndf}=31/28$) and  
$\alpha_p (D^0 \to \pi^- \ell^+ \nu) = 0.10 \pm 0.21_\mathrm{stat}  \pm 0.10_\mathrm{syst}$ ($\chi^2/\mbox{ndf}=6.4/10$).

The fitted values for $f_+^{K,\pi}(0)$ vary little for the different fits, for the modified pole model the results are
$f_+^K(0) = 0.695 \pm 0.007_{\mathrm{stat}} \pm 0.022_{\mathrm{syst}}$ and $f_+^\pi(0) = 0.624 \pm 0.020_{\mathrm{stat}} \pm 0.030_{\mathrm{syst}}$.

The measured form factors $f_+^{K,\pi}(q^2)$ are shown in Figure \ref{fig:X} with predictions
of the simple pole model, unquenched~\cite{ref:unquenched2} and quenched~\cite{ref:quenched} LQCD.  To
obtain a continuous curve for $f_+$ from the LQCD values reported  at
discrete $q^2$ points, the values were fitted by a parabola, which is
found to fit well within the stated theoretical errors and is not
associated with any specific model.  To quantify the degree of
agreement, a $\chi^2$/ndf is calculated between this measurement and the interpolated LQCD curve within the $q^2$ range for which LQCD predictions are made.  For the kaon modes,
$\chi^2/\mbox{ndf}$ is $28/18$ ($34/23$), for the pion modes $9.8/5$ ($3.4/5$); correlations induced by the fit of the calculated $q^2$ points to a parabola have been considered.

\subsection{$D \to K e \nu_e$ at \babar} 

{\small This subsection is an adapted excerpt of \babar's publication~\cite{babarkenu}.}

\vspace{3mm}
The corresponding \babar analysis~\cite{babarkenu} is using  
a total integrated luminosity of $75$ fb$^{-1}$ collected
during the years 2000-2002. 
It measures the $q^2$ variation and the absolute
value of the hadronic form factor at $q^2 = 0$ for the
decay $D^0 \rightarrow K^- e^+ \nu_e(\gamma)$. Normalizing to $D^0 \to K^-\pi^+$, it also gives a value for its branching fraction.
In contrast to the Belle analysis, a semi-inclusive reconstruction technique is used 
to select semileptonic decays with less resolution, but much higher efficiency. As a result of this approach,
events with a photon radiated during the $D^0$ decay are included in the signal.  

$D^0 \rightarrow K^- e^+ \nu_e (\gamma)$  decays are reconstructed in 
$e^+e^- \rightarrow c\bar{c}$  events where the $D^0$ originates from
the $D^{*+} \rightarrow D^0 \pi^+$.
Charged and neutral particles are boosted to the center of mass system (c.m.) 
and the event thrust
axis is determined.
The direction of this axis is required to be in the interval
$|\cos(\theta_{{\rm thrust}})|<0.6$ to minimize the loss of particles in 
regions close to the beam axis.
A plane perpendicular to the thrust axis is used to define two hemispheres,
equivalent to the two jets produced by quark fragmentation. In each hemisphere, 
pairs of oppositely charged leptons and kaons are searched for.
For the charged lepton candidates 
only electrons or positrons with c.m. momentum
greater than 0.5~GeV/$c$ are considered.
 
Since the $\nu_e$ momentum is unmeasured, a kinematic fit is performed, constraining the invariant mass of the candidate
$e^+K^-\nu_e$ system to the $D^0$ mass. In this fit,
the $D^0$ momentum and the neutrino energy are estimated from the other particles 
measured in the event. The $D^0$ direction is taken as the direction opposite
to the sum of the momenta of all reconstructed particles in the event, except for the kaon and the
positron associated with the signal candidate. 
The energy of the jet is determined from the total c.m. energy and from the measured
masses of the two jets.
The 
neutrino energy is estimated as the difference between the total energy of the jet
and the sum of the energies of all reconstructed particles in the hemisphere. 
A correction, which depends on the value of the missing energy measured in the opposite jet,
is applied to account for the presence of missing energy due to 
particles escaping detection, even in the absence of a neutrino from the $D^0$ decay.

\begin{table*}[t]
  \caption {\babar: Fitted values of the parameters corresponding to 
different parameterizations of $f_+(q^2)$. The last column
gives the $\chi^2/NDF$ of the fit when using the value expected for the
parameter.}

\begin{center}
  \begin{tabular}{lclcl}
    \hline\hline
Theoretical &Unit & Parameters&$\chi^2$/NDF& Expectations \\
ansatz & & & & [$\chi^2$/NDF]\\
\hline
$z$ expansion& & $a_1  = -2.5 \pm 0.2 \pm 0.2$ &5.9/7 &  \\
   &                    & $a_2  = 0.6 \pm 6. \pm 5.$& &\\

Modified pole &&$\alpha_{\rm pole} = 0.377 \pm 0.023 \pm 0.029$ & 6.0/8& \\

Simple pole &$\rm{GeV/c^2}$ &$m_{\rm pole} = 1.884 \pm 0.012 \pm 0.015$&7.4/8& $2.112~[243/9]$\\

\hline \hline
  \end{tabular}
\end{center}
  \label{tab:fittedparam}
\end{table*}

Background events arise from $\Upsilon(4S)$ decays and hadronic 
events from the continuum. 
To  reduce the contribution from $B\bar{B}$ events, selection criteria exploiting the topological differences to
events with $c\bar{c}$ fragmentation are used.
Background events from the continuum arise mainly from charm particles
Because charm hadrons take a large
fraction of the charm quark energy, charm decay products
have higher average energies and different angular distributions  (relative to 
the thrust axis or to the $D$ direction) compared with
other particles in the hemisphere emitted from the hadronization
of the $c$ and $\overline{c}$ quarks. Selection criteria based on these considerations are applied to suppress this kind of background.

The remaining background from $c\bar{c}$-events can be divided into peaking (60$\%$) and 
non-peaking (40$\%$) candidates. 
Peaking events are those background events whose distribution is peaked around the signal region.
These are mainly
events with a real $D^{*+}$ in which the slow $\pi^+$ is
included in the candidate track combination. 
Backgrounds from $e^+e^-$ annihilations into light $u\bar{u},~d\bar{d},~s\bar{s}$ quarks and $B\bar{B}$ events
are non-peaking.
To improve the accuracy of the reconstructed $D^0$ momentum, 
the nominal $D^{*+}$ mass 
is added as a constraint in the previous fit and only events with
a $\chi^2$ probability higher than 1$\%$ are kept,
resulting in 85260 selected $D^0$ candidates containing an estimated number
of 11280 background events.
The non-peaking component comprises $54\%$ of the background.
Detailed studies have been performed to understand corrections (and the connected systematics) of various components of the peaking background.

To obtain the true $q^2$ distribution, the measured one
has to be corrected for selection efficiency
and detector resolution effects. This is done using an unfolding algorithm
based on MC simulation of these effects:  
Using Singular Value Decomposition (SVD)
\cite{ref:svd} of the resolution matrix, the unfolded $q^2$ distribution for signal events, 
corrected for resolution 
and acceptance effects, is obtained.
This approach provides the full covariance matrix for the
bin contents of the unfolded distribution.
To verify that the $q^2$ variation of the selection 
efficiency is well described by the simulation, a control sample of 
$D^0 \rightarrow K^- \pi^+ \pi^0$ is reconstructed
as if they were $K^- e^+ \nu_e$ events, and indicates no significant bias.
With a second control sample of $D^0 \rightarrow K^- \pi^+$, the
accuracy of the $D^0$ direction and missing energy reconstruction
for the $D^0 \rightarrow K^- e^+ \nu_e$ analysis is checked.
This information is used in
the mass-constrained fits and thus influences the $q^2$ reconstruction.
Once the simulation is tuned to reproduce the results obtained on data
for these parameters, the $q^2$ resolution distributions agree
very well.
One half of the measured variation on the fitted parameters from these corrections 
has been taken as a systematic uncertainty.

Effects from a momentum-dependent difference between data
and simulated events on
the charged lepton and on the kaon
identification have been found to be $<2\%$ and included in the corrections and systematics.
Care has also be taken to correctly understand 
radiative decays where 
$q^2=(p_D-p_K)^2 =(p_e+p_{\nu}+p_{\gamma})^2$.
Corresponding corrections have been applied and the corresponding uncertainties
enter in the systematic uncertainty evaluation.
Toy simulations have been used to verify that the statistical 
precision obtained for each binned unfolded value is correct and if 
biases generated by removing information  are under control.

The fit to a model is done by
comparing the number of events measured in a given bin of $q^2$
with the expectation from the exact analytic integration
of the expression $\left | \vec{p}_K (q^2) \right |^3 \left |f_+(q^2) \right |^2$
over the bin range, with the overall normalization left free. 
The summary of the fits to the normalized $q^2$ distributions is presented in Table \ref{tab:fittedparam}.
As long as the form factor parameters are left free in the fit, the fitted 
distributions agree well with the data and it is not possible to reject any of the parameterizations.  
As also observed by Belle and other experiments, the simple pole model
ansatz, with $m_{\rm pole}=m_{D_s^*}=2.112~\rm{GeV/c^2}$ does not reproduce
the measurements. 

\begin{figure}[!htb]
\begin{center}
\includegraphics[width=8cm]{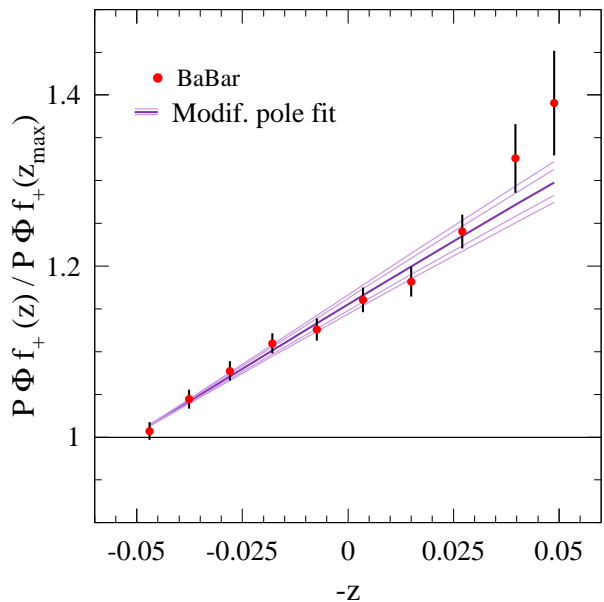}
\caption{ \babar: Measured values
for $P \times \Phi \times f_+$ are plotted versus $-z$ and 
requiring that $P \times \Phi \times f_+ = 1$ for $z=z_{\rm max}$.
The straight lines represent the result for the modified pole ansatz, 
the fit in the center and the statistical and total uncertainty.}
\label{fig:zhill}
\end{center}
\end{figure}

Figure \ref{fig:zhill} shows the product $P \times \Phi \times f_+$ as a function of $z(q^2)$ (defined above), where $P \times \Phi$ is the theoretical normalization~\cite{ref:hill1}, which constrains this product to unity at $z=z_{\rm max}$ (equivalent to $q^2=0$). 
The data are compatible with a linear dependence, which is fully consistent with the modified pole ansatz for $f_+(q^2)$.

The $D^0 \rightarrow K^- e^+ \nu_e$ branching fraction
is measured relative to the reference decay channel,
$D^0 \rightarrow K^- \pi^+$:
\begin{equation}
R_D = \frac{BR(D^0 \rightarrow K^- e^+ \nu_e)_{\rm data}}{BR(D^0 \rightarrow K^- \pi^+)_{\rm data}} 
\end{equation}
Using slightly adapted selection criteria,
after background subtraction there remain $76283 \pm 323$ 
events of $D^0 \rightarrow K^- e^+ \nu_e$ in data.
To select $D^0 \rightarrow K^- \pi^+ $ candidates, care has been taken to do this in the most similar way possible.
After background subtraction and the necessary corrections,
there are $134537\pm 374$ candidates 
selected in the interval 
$\delta(m)\in [0.142,~0.149]~\rm{GeV/c^2}$.

A summary of the systematic uncertainties on $R_D$ is given in~\cite{babarkenu}.
The measured relative decay rate is:

\begin{equation}
R_D = 0.9269 \pm 0.0072 \pm 0.0119.
\end{equation}
\noindent
Using the world average for the branching fraction
$BR(D^0 \rightarrow K^- \pi^+) =(3.80 \pm 0.07) \%$
\cite{ref:pdg06}, gives
$BR(D^0 \rightarrow K^- e^+ \nu_e(\gamma)) = (3.522 \pm 0.027 \pm 0.045 \pm 0.065)\%$,
where the last quoted uncertainty corresponds to the 
accuracy on $BR(D^0 \rightarrow K^- \pi^+)$.

The value of the hadronic form
factor at $q^2=0$ can be obtained as
\begin{equation}
f_+(0)=\frac{1}{\left | V_{cs} \right |}
\sqrt{\frac{24 \pi^3}{G_F^2}\frac{BR}{\tau_{D^0} I}},
\end{equation}
\noindent where
$BR$ is the measured $D^0 \rightarrow K^- e^+ \nu_e$ branching fraction, 
$\tau_{D^0}=(410.1\pm1.5)\times10^{-15}~s$ \cite{ref:pdg06} is the $D^0$ lifetime and
$I=\int_0^{q^2_{\rm max}}{\left | \vec{p}_K (q^2) \right |^3 \left |f_+(q^2)/f_+(0) \right |^2}~dq^2$.
To account for the variation of the form factor
within one bin, and in particular to extrapolate the result at $q^2=0$, the pole
mass and the modified pole ans\"atze have been used; the corresponding values obtained for 
$f_+(0)$ differ by 0.002. Taking the average between these two values and including
their difference in the systematic uncertainty,
this gives
\begin{equation}
f_+(0) = 0.727 \pm 0.007 \pm 0.005 \pm 0.007, 
\end{equation}
where the last quoted uncertainty corresponds to the accuracy
on $BR(D^0 \rightarrow K^- \pi^+)$, $\tau_{D^0}$ and $\left | V_{cs} \right |$.
For the $z$ expansion, this corresponds to  
$a_0=(2.98 \pm 0.01 \pm0.03 \pm0.03)\times 10^{-2}$.

\section{Semileptonic Decays to Vector Mesons}

\begin{figure}[htbp]
  \begin{center}
    \includegraphics[width=8cm]{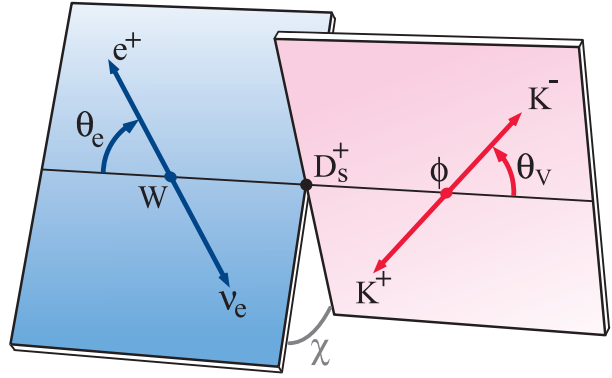}
  \end{center}
  \caption[]{ Definition of the angles $\theta_e$, $\theta_V$, and $\chi$.}
   \label{fig:dsphienu_decay}
\end{figure}

The differential semileptonic decay rate of a scalar meson to a vector meson, specifically, $\Ds \rightarrow \phi e^+ \nu_e$,  
depends on the four variables $q^2$, $\theta_e$, $\theta_V$ and $\chi$~\cite{ref:KS}, depicted in Fig. \ref{fig:dsphienu_decay}.

Neglecting the electron mass, the differential decay rate as function of these four variables depends in a given way~\cite{babarphienu,ref:ks2}
on three form factors
\begin{eqnarray}
A_{1,2}(\qsq) & = & {A_{1,2}(0)\over 1-\qsq/m_A^2} \\
V(\qsq) & = & {V(0)\over 1-\qsq/m_V^2}
\label{eq:pole}
\end{eqnarray}
with the pole masses $m_A = 2.5~\GeVcd$ and $m_V = 2.1~\GeVcd$. 
Measurements have usually been expressed in terms of the ratios
of the form factors at $\qsq=0$, namely:
\beq
r_V~=~V(0)/A_1(0)~~{\rm and}~~r_2~=~A_2(0)/A_1(0).
\eeq

Based on a prediction by~\cite{ref:fk}, $r_V$ is a constant depending only on particle masses,
\beq
r_V~=~\frac{\left ( m_{D_s}+m_{\phi}\right )^2}{m_{D_s}^2+m_{\phi}^2}~=~1.8.
\label{eq:svetla}
\eeq

\subsection{$D_s \to \phi e \nu_e$ at \babar}

{\small This subsection is an adapted excerpt of \babar's publication~\cite{babarphienu}.}

\vspace{3mm}
BaBar has presented a study of the hadronic form factors 
for the vector meson decay $\Ds \rightarrow \phi e^+ \nu_e$ with $\phi \to K^+K^-$~\cite{babarphienu} (results still preliminary).
This analysis is based on a fraction of the total available \babar\ data sample, 
corresponding to integrated luminosities of $78.5~\fb^{-1}$ recorded on the $\FourS$ resonance.
It focuses on semileptonic decays of $\Ds$ mesons which are produced via $e^+ e^- \to c\overline{c}$ annihilation.  $D_s$ mesons produced in $\BB$ events are not included and treated as background. 
  
Similar to \babar's $D^0 \to Ke\nu_e$ analysis presented above, a plane perpendicular to the thrust axis is used to define two hemispheres, 
equivalent to the two jets produced by quark fragmentation.
In each hemisphere,  decay products of the $\Ds$, a charged lepton and two oppositely charged kaons are searched for. Charged leptons 
are required to have a c.m. momentum larger than 0.5~$\GeVc$.
The unmeasured neutrino momentum is determined in a way similar to the  $D^0 \to Ke\nu_e$ analysis presented above.

\begin{figure}[t]
  \begin{center}
   \includegraphics[width=8cm]{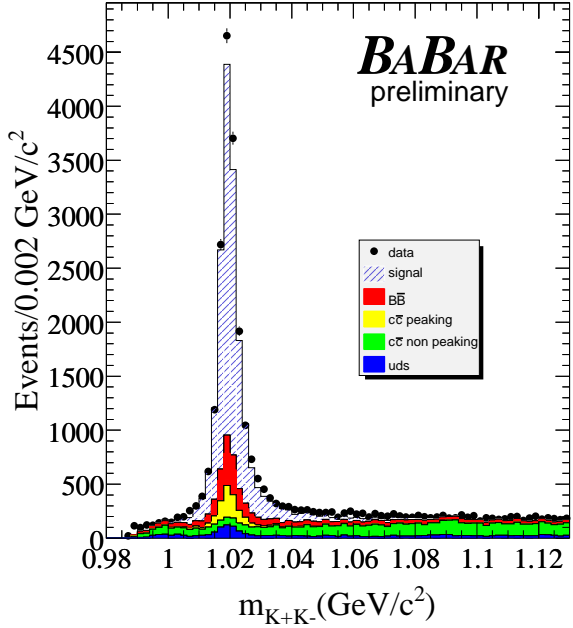}
  \end{center}
  \caption[]{\babar: $K^+K^-$ invariant mass distribution from data and simulated events. MC events have been normalized to the data luminosity according to the different cross sections. The excess of signal events in the $\phi$ region
can be attributed to a different production rate and decay branching fraction
of $\Ds$ mesons in data and in simulated events. Dedicated studies have been
done to evaluate the amount of peaking background in real events.}
   \label{fig:bkg_comp}
\end{figure}

Figure~\ref{fig:bkg_comp} shows the $K^+ K^-$ invariant mass distribution for the selected decays compared to MC simulation and the composition of the background. 
$\phi$~candidates are defined as $K^+ K^-$ pairs with an invariant mass 
in the interval from 1.01 and 1.03~$\GeVcd$.
Various selection criteria are applied to suppress the background~\cite{babarphienu}.
About 71$\%$ of the total background include a true $\phi$ decay combined 
with an electron from another source, namely $B$ meson decays (41$\%$), charm particle decays (25$\%$), photon conversions or Dalitz decays (24$\%$), and the rest are fake electrons.
These $\phi$ mesons are expected to originate from the 
primary vertex, or from a secondary charm decay vertex. 

\begin{figure}[t]
  \begin{center}
   \includegraphics[width=8cm]{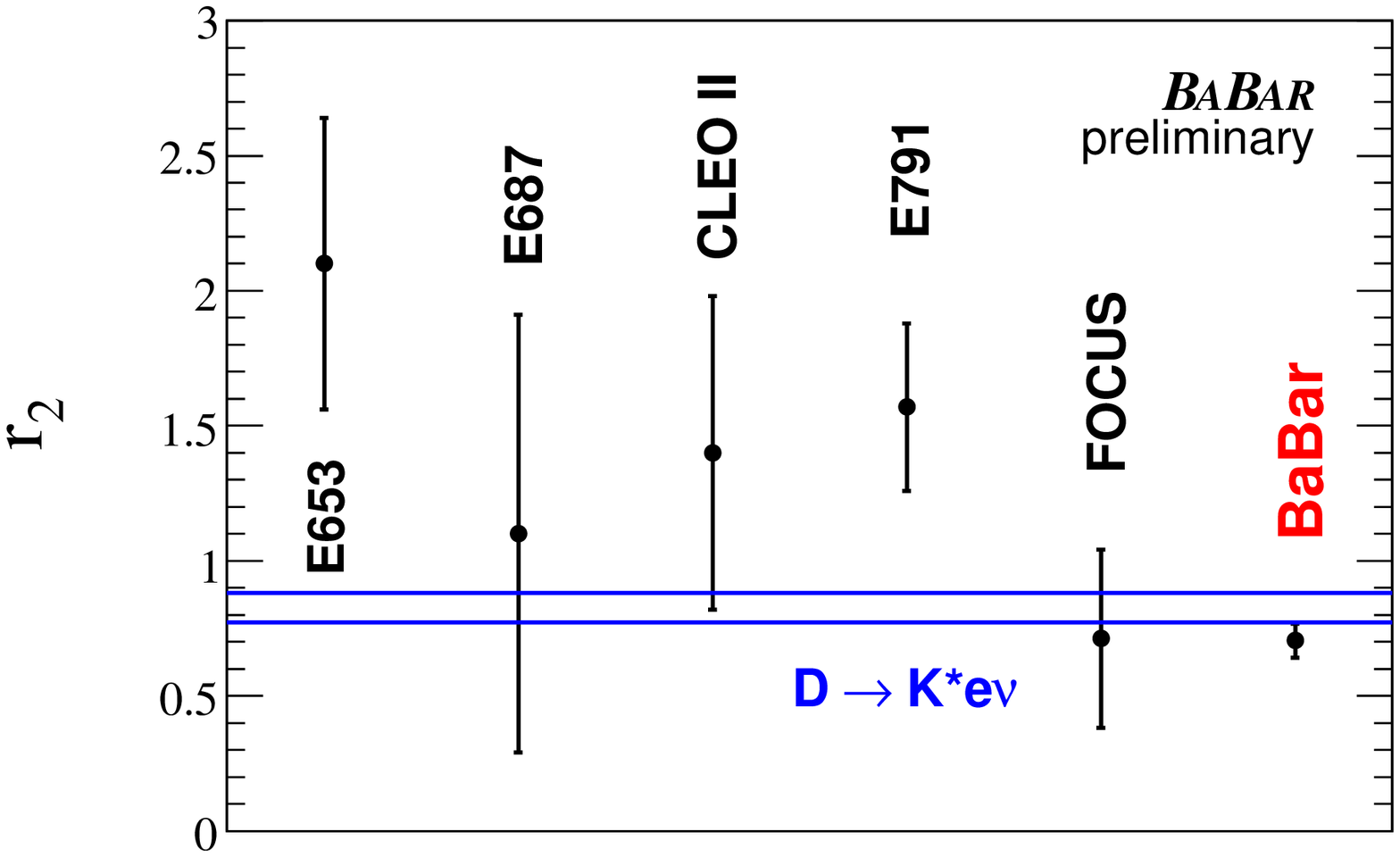}
   \includegraphics[width=8cm]{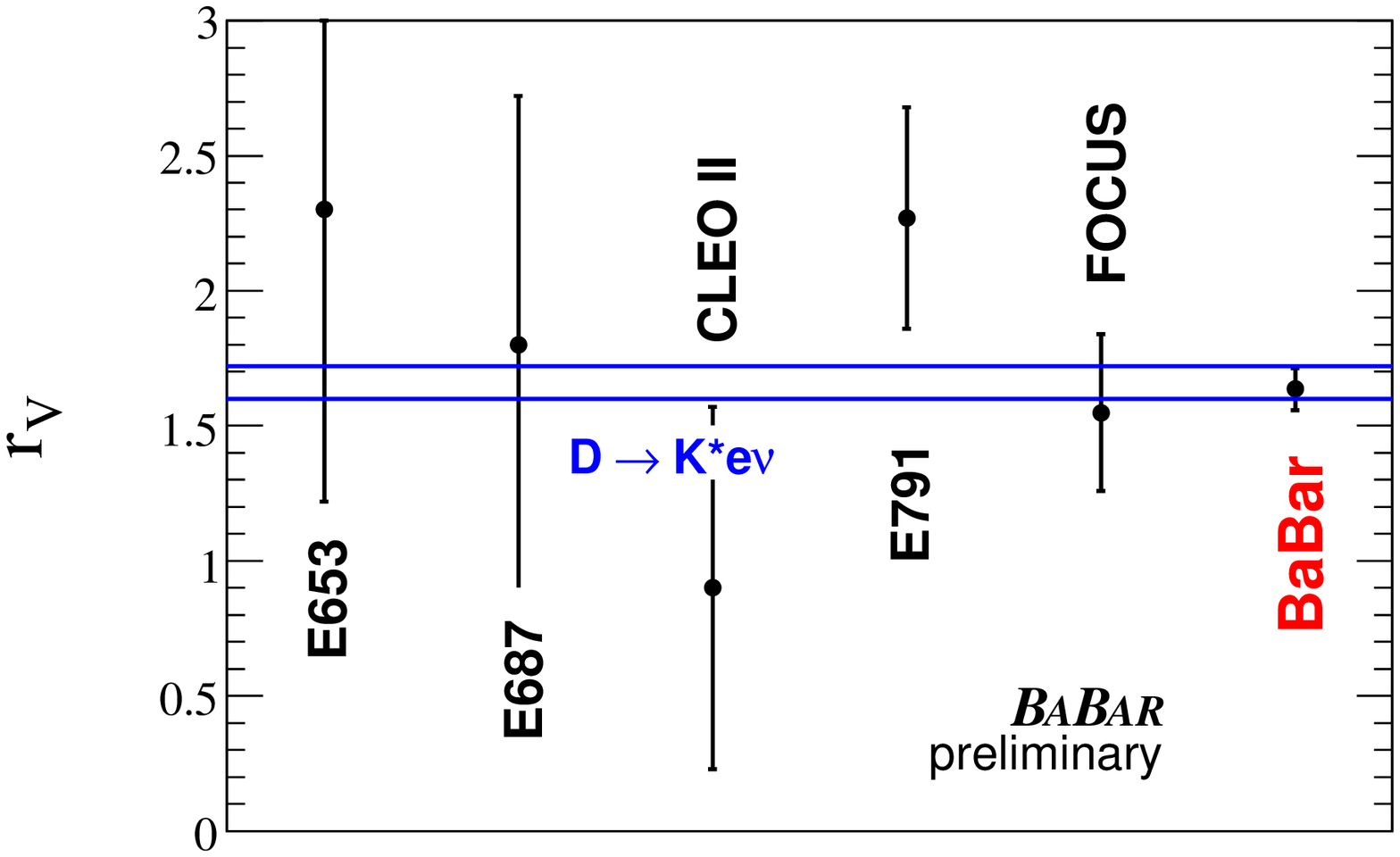}
  \end{center}
  \caption[]{\babar: Results from previous experiments and present measurement
of $r_2$ and $r_V$ in $\Ds \rightarrow \phi e^+ \nu_e $ decays. The error bars represent the statistical and systematic uncertainties added in quadrature.
These measurements for $D_s \to \phi e^+ \nu_e $ decays are compared with the average of similar measurements obtained for  $D \rightarrow K^* e^+ \nu_e $ decays. The $\pm$ one sigma range is indicated by the two parallel lines. 
}
   \label{fig:r2rv}
\end{figure}

A maximum likelihood fit is performed to the four-dimensional decay distribution in the reconstructed
variables $q^2_r$, 
$\cos(\theta_V)_r$, $\cos(\theta_e)_r$ and $\chi_r$
using the likelihood function
\beq
{\cal L} = - \sum_{i=1}^{625} \ln {{\cal P}(n^{\rm data}_i |n^{\rm MC}_i) }.
\eeq
\noindent
In this expression, for each bin $i$, ${\cal P}(n^{\rm data}_i |n^{\rm MC}_i)$ is the
Poisson probability to observe $n^{\rm data}_i$ events, when
$n^{\rm MC}_i$ are expected.
Considering the typical resolutions and the available statistics, 
the four variables are divided into 4 bins each, corresponding to a four-dimensional array with a total number of bins of $625$.

To determine the expected number of signal events, a dedicated sample of signal events is generated in MC with a uniform decay phase space distribution, and each event
is weighted using the differential
decay rate divided by $p_{\phi}$.
Two of the four variables, $\cos(\theta_V)$ and $\chi$, are integrated (averaged),
taking advantage of the fact that the estimated background rate is flat in these variables.
The background components are normalized to correspond to the 
expected rates for the integrated luminosity of the data sample.
The absolute normalization for signal events $(N_S)$ is left free to vary
in the fit. In each bin $(i)$, the expected number of events is evaluated 
to be:
\beq
n^{\rm MC}_i = N_S \frac{\sum_{j=1}^{n_i^{\rm signal}} w_j(\lambda_k)}{W_{\rm tot}(\lambda_k)}~+~ n^{\rm bckg.}_i.
\eeq
Here $n_i^{\rm signal}$ refers to the number of simulated signal events, with reconstructed values of the four variables corresponding to bin $i$. The weight $w_j$ is evaluated for each event, using the generated values of the 
kinematic variables, thus accounting for resolution effects.
$W_{\rm tot}(\lambda_k)=\sum_{j=1}^{N^{\rm signal}} w_j(\lambda_k)$
is the sum of the weights for all simulated signal events which have been 
generated according to a uniform phase space distribution. $N_S$ and 
$\lambda_k$ are the parameters to be fitted. Specifically, the free parameters   $\lambda_k$ are $r_V$, $r_2$, and parameters which define
$q^2$ dependence of the form factors. To avoid having to introduce finite
ranges for the fit to the pole masses, $m_i$, we define $m_i = 1 +\lambda_i^2$. This expression ensures that $m_i$ is always larger than $q^2_{max.}\simeq 0.9$ $\GeV^2$. 

The fit to the four-dimensional data distribution is performed using simulated signal events generated according to a uniform phase space distribution.
Signal MC events are weighted to correct for differences 
in the quark fragmentation process between data and simulated events. 

Differences between data and MC have been measured 
using $\Ds \rightarrow \phi \pipl$ decays, according corrections were applied.
The influence of combinatorial background has been studied and considered in the systematics.
Background from $\phi$ mesons produced in $D$ or $B$ decays results in a further correction in MC (to calibrate the $\phi$ production rate) and corresponding systematics.
The effect of uncertainties due to finite MC statistics and background estimation on the fit has been studied with toy simulations, and was also included in the final results.
Remaining detector effects have been determined with control data samples.
Using $\Dstarp \rightarrow \Do \pi^+$ and $~\Do  \rightarrow K^-\pi^+\pi^0$
events it has been verified that differences between data and
simulated events in the resolution of the variables $\qsq$
and $\cos(\thl)$ are small compared with other
sources of systematic uncertainties. They have been neglected at present.

Using fixed values for the pole masses ($m_A =
2.5~\GeVcd$ and $m_V = 2.1~\GeVcd$), the final fit results including all corrections are:
\begin{eqnarray}
N_S & = & 12886 \pm 129 \nonumber\\r_V & = & 1.636\pm0.067\pm0.038 \nonumber\\r_2 & = & 0.705\pm 0.056\pm 0.029.\nonumber
\end{eqnarray}
\noindent
Keeping $m_V$ fixed, for which there is no sensitivity, and adding $m_A$ as additional free parameter, the fit results in
\begin{eqnarray}
N_S & = & 12887 \pm 129 \nonumber\\r_V & = & 1.633\pm0.081\pm0.068 \nonumber\\r_2 & = & 0.711\pm 0.111\pm 0.096 \nonumber\\
m_A & = & 2.53^{+0.54}_{-0.35}\pm0.54~\GeVcd.\nonumber
\end{eqnarray}

The measurements of the parameters $r_V$ and $r_2$ for the semileptonic
decay $\Ds \rightarrow \phi e^+ \nu_e$ have an accuracy similar
to the one obtained for $D \rightarrow K^* e^+ \nu_e $ decays \cite{ref:wiss},
see Fig. \ref{fig:r2rv}.

\section{Leptonic Decays}

The purely leptonic decay $D_s^+ \to \ell^+ \nu_\ell$ (the charge conjugate mode is implied throughout this paper) is theoretically a rather clean decay; in the Standard Model (SM), the decay is mediated by a single virtual $W^+$-boson. The decay rate is given by
\begin{equation} \label{eqn:gammunu}
	\Gamma(D_s^+ \to \ell^+ \nu_\ell) = \frac{G^2_F}{8\pi}f^2_{D^+_s}m^2_\ell M_{D^+_s} \left( 1-\frac{m^2_\ell}{M^2_{D^+_s}} \right)^2 |V_{cs}|^2,
\end{equation}
where $G_F$ is the Fermi coupling constant, $m_\ell$ and $M_{D^+_s}$ are the masses of the lepton and of the $D_s$ meson, respectively. $V_{cs}$ is the corresponding CKM-matrix element, while all effects of strong interaction are accounted for in the decay constant $f_{D^+_s}$. Due to helicity conservation, the decay rate is highly suppressed for electrons. Since the detection of $\tau$'s involve additional neutrinos, the muon mode is experimentally the cleanest and most accessible mode.

\subsection{$D_s \to \mu \nu_{\mu}$ at \babar}

{\small This subsection is an adapted excerpt of \babar's publication~\cite{babardsmunu}.} 

\vspace{3mm}
\babar performed a measurement of the ratio
$\Gamma(D_s \to \mu\nu_\mu)/\Gamma(D_s \to \phi\pi)$ and the decay
constant $f_{D_s}$, based on a total integrated luminosity of $230.2$ fb$^{-1}$ . 

In order to measure
$\Ds\to\mu^+\nu_\mu$, the decay chain $\Dss\to\gamma\Ds,
\Ds\to\mu^+\nu_\mu$ is reconstructed from \Dss mesons produced in the
hard fragmentation of continuum \ccbar events.  The subsequent decay
results in a photon, a high-momentum \Ds and daughter muon and
neutrino, lying mostly in the same hemisphere of the event.  Signal
candidates are required to lie in the recoil of a fully reconstructed
\Dz, \Dp, \Ds, or \Dstarp meson (the ``tag'') reconstructed in a variety of modes~\cite{babardsmunu} wherein the tag flavor is uniquely determined.
To eliminate signal from \B decays, the minimum tag momentum is chosen
to be close to the kinematic limit for charm mesons arising from \B
decays.

For each event a single tag candidate is chosen and then used in the
subsequent analysis.  To pick this tag among multiple candidates
within an event (there are 1.2 candidates on average in events with at
least one candidate) modes of higher purity are preferred. In events
where two tag candidates are reconstructed in the same mode, the
quality of the vertex fit of the $D$ meson is used as a secondary
criterion. After subtracting combinatorial background there are
$5*10^5$ charm tagged events with a muon amongst the recoiling
particles.

The signature of the decay $\Dss\to\gamma\Ds$ is a narrow peak in the
distribution of the mass difference $\Delta M =
M(\mu\nu\gamma)-M(\mu\nu)$ at \unit[143.5]{\mevcc}. The \Dss signal is
reconstructed from a muon and a photon candidate in the recoil of the
tag. Muons are identified as non-showering tracks penetrating the IFR.
The muon must have a momentum of at least \unit[1.2]{\gevc} in the
center-of-mass (CM) frame and have a charge consistent with the tag
flavor. 
Clusters of energy in the EMC not
associated with charged tracks and exceeding an CM energy of $\unit[0.115]{\gev}$ are identified as photon candidates.

The CM missing energy ($E^*_{\rm{miss}}$) and momentum ($\vec
p^*_{\text{miss}}$) are calculated from the four-momenta of the incoming
\epem, the tag four-momentum, and the four-momenta of all remaining
tracks and photons in the event. The energy of the charged particles
that do not belong to the tag is calculated from the track momentum
under a pion mass hypothesis. Assigning a mass according to the most
likely particle hypothesis has negligible effect on the missing energy
resolution. 

The neutrino CM four-momentum ($p^*_\nu=(|\vec p^*_\nu|,\vec
p^*_\nu)$) is estimated from the muon CM four-momentum ($p^*_\mu$) and
$\vec p^*_{\text{miss}}$, using a technique adopted from
Ref.~\cite{Chadha:1998zh}. The difference $|\vec p^*_{\rm miss} - \vec
p^*_\nu|$ is minimized, while the invariant mass of the neutrino-muon
pair is required to be the known mass of the \Ds
\cite{Eidelman:2004wy}.
The muon CM
four-momentum ($p^*_\mu$) is combined with $p^*_\nu$ to form the \Ds
candidate.  
The \Ds candidate is then combined
with a photon candidate to form the \Dss.  
The selection requirements on $E^*_{\text{miss}}$, 
$p_{\text{corr}}$, and other variables have been optimized
to maximize the significance $s/\sqrt{s+b}$, where
$s$ and $b$ are the signal and background yields expected in the data
set. 

One class of background are events $\epem\to f\bar f$, where $f =
u,d,s,b$, or $\tau$, which do not contain a real charm tag. The
contribution of these events is estimated from data using the tag
sidebands. In addition there are events $\epem\to\ccbar$ where the tag
is incorrectly reconstructed. Although these events potentially
contain the signal decay, they are also subtracted using the tag
sidebands.  These two sources amount to $\unit[\approx42]{\%}$ of the
background.
A second class of background events ($\unit[\approx26]{\%}$) are
correctly tagged $\ccbar$ events with the recoil muon coming from a
semileptonic charm decay or from $\tau^+\to\mu^+\nu_\mu\bar\nu_\tau$.
This includes events $\Dss\to\gamma\Ds\to\gamma\tau^+\nu_\tau,\ 
\tau^+\to\mu^+\nu_\mu\bar\nu_\tau$. To estimate the size and shape of
this background contribution, the analysis is repeated, substituting a
well-identified electron for the muon. Except for a small phase-space
correction, the widths of weak charm decays into muons and electrons
are assumed to be equal. QED effects such as bremsstrahlung
($e^+\to\gamma e^+$) energy losses and photon conversion ($\gamma\to
e^+e^-$), where the muon equivalents have a much lower rate, are
explicitly removed.  In particular, bremsstrahlung photons found in
the vicinity of an electron track are combined with the track.  The
small number of events with an electron from a converted photon that
survive the selection are suppressed by a photon conversion veto,
using the vertex and the known radial distribution of the material in the
detector.
The muon selection efficiency as a function of momentum and direction is
measured using $\epem\to\mu^+\mu^-\gamma$ events, while radiative
Bhabha events are used to quantify the electron efficiency. The ratio
of muon to electron efficiencies is applied as a weight to each
electron event.
The remaining backgrounds are estimated from simulation.

\begin{figure}[tbp]
  \centering
  \includegraphics[width=8cm]{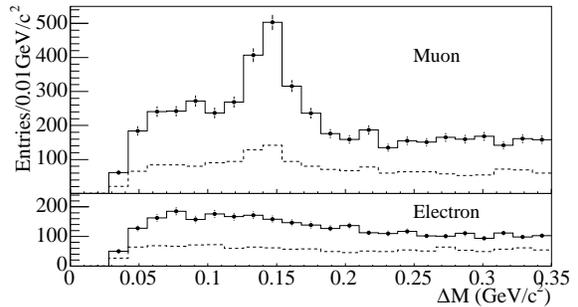}
  \caption{\babar: $\Delta M$ distribution of charm-tagged events passing the signal
    selection.  The tag can be from the tag signal region (solid
    lines) or the sidebands (dashed lines). In the bottom plot the
    signal muon is replaced with an electron to estimate the
    semileptonic charm and $\tau$ decay background.
 }
  \label{fig:datafour}
\end{figure}

\begin{figure}[t]
  \centering
  \includegraphics[width=8cm]{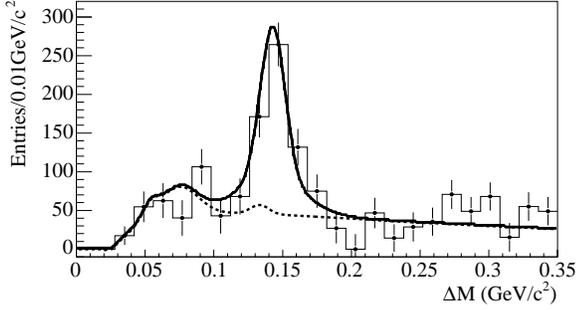}
  \caption{\babar: $\Delta M$ distribution after the tag sidebands and
    the electron sample are subtracted. The solid line is the fitted
    signal and background distribution $(N_{\text{Sig}}f_{\text{Sig}} +
    N_{\text{Bkgd}}f_{\text{Bkgd}})$, the dashed line is the background
    distribution $(N_{\text{Bkgd}}f_{\text{Bkgd}})$ alone. }
  \label{fig:data}
\end{figure}

Events that pass the signal selection are grouped into four sets,
depending on whether the tag lies in the signal region or the sideband
regions, and on whether the lepton is a muon or an electron
(Fig.~\ref{fig:datafour}).  For each lepton type the sideband $\Delta
M$ distribution is subtracted.  The electron distribution, scaled by
the relative phase-space factor (0.97) appropriate to semileptonic
charm meson decays and leptonic $\tau$ decays is then subtracted from
the muon distribution.  The resulting $\Delta M$ distribution is
fitted with a function $(N_{\text{Sig}}f_{\text{Sig}} +
N_{\text{Bkgd}}f_{\text{Bkgd}})(\Delta M)$, where $f_{\text{Sig}}$ and
$f_{\text{Bkgd}}$ describe the simulated signal and background $\Delta
M$ distributions. The function $f_{\text{Sig}}$ is a double Gaussian
distribution. The function $f_{\text{Bkgd}}$ consists of a double and a
single Gaussian distribution describing the two peaking background
components, and a function~\cite{babardsmunu} describing the flat
background component. The relative sizes of the background components,
along with all parameters except $N_{\text{Sig}}$ and $N_{\text{Bkgd}}$
are fixed to the values estimated from simulation. The $\chi^2$ fit
yields $N_{\text{Sig}} = \SigFitYield\pm\SigFitYieldError(stat)$ signal
events and has a fit probability of \unit[\SigFitYieldGOF]{\%}
(Fig.~\ref{fig:data}).

The branching fraction of \dstomunu cannot be determined directly,
since the production rate of $D_s^{(*)+}$ mesons in \ccbar
fragmentation is unknown.  Instead the partial width ratio
$\Gamma(\dstomunu)/\Gamma(\dstophipi)$ is measured by reconstructing
\dsstodstophipi decays. The \dstomunu branching fraction is evaluated
using the measured branching fraction for \dstophipi.

\begin{figure}[t]
  \centering
  \includegraphics[width=8cm]{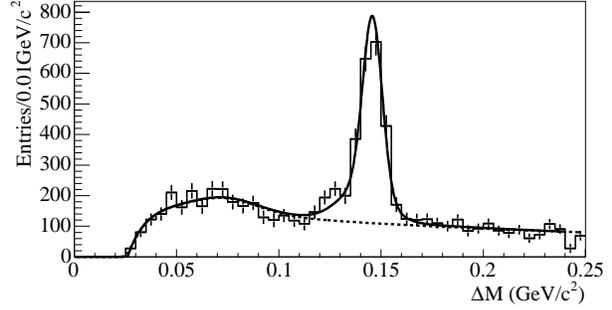}
  \caption{\babar: $\Delta M$ distribution of selected \dsstodstophipi events
    after the tag sideband is subtracted. The solid line is the fitted
    signal and background distribution
    $(N_{\phi\pi}f_{\phi\pi} + N_{\phi\pi\text{Bkgd}}f_{\phi\pi \text{Bkgd}})$, the dashed line is the
    background distribution $(N_{\phi\pi \text{Bkgd}} f_{\phi\pi \text{Bkgd}})$ alone. }
  \label{fig:dsphipi}
\end{figure}

Candidate $\phi$ mesons are reconstructed from two kaons of opposite
charge. The $\phi$ candidates are combined with charged pions to form
\Ds meson candidates. Both times a geometrically constrained fit is
employed, and a minimum requirement on the fit quality is made. The
$\phi$ and the \Ds candidate masses must lie within $2\,\sigma$ of
their nominal values, obtained from fits to simulated events and data.
Photon candidates are then combined with the \Ds to form \Dss
candidates. The same requirements on the CM photon energy and \Dss
momentum as in the \dstomunu signal selection are made. 
Data events that
pass the selection are grouped into two sets: the tag signal and
sideband regions. After the tag sideband has been subtracted from the
tag signal $\Delta M$ distribution, the remaining distribution is
fitted with $(N_{\phi\pi}f_{\phi\pi}+N_{\phi\pi \text{Bkgd}}f_{\phi\pi
  \text{Bkgd}})(\Delta M)$, where $f_{\phi\pi}$ is a triple Gaussian,
describing the simulated \dsstodstophipi signal, and $f_{\phi\pi
  \text{Bkgd}}$ consists of a broad Gaussian centered at $70\,\mevcc$
and a function~\cite{babardsmunu} describing the simulated background
$\Delta M$ distributions. The Gaussian describes the background
$\Dss\to\piz\Ds\to\piz\phi\pip$ where the photon candidate originates
from the \piz. The relative sizes of the background components, along
with all parameters except $N_{\phi\pi}$, $N_{\phi\pi\text{Bkgd}}$,
and the mean of the peak are fixed to the values estimated from
simulation.  The $\chi^2$ fit yields
$N_{\phi\pi}=\dsPhiPiDataFitYield\pm\dsPhiPiDataFitYieldError$ events
and has a probability of \unit[\dsPhiPiDataGOF]{\%}
(Fig.~\ref{fig:dsphipi}). From simulation $48\pm23$ events
$\Dss\to\gamma\Ds\to\gamma f_0(980)(\Kp\Km)\pip$ are expected to
contribute to the signal, where the error is mostly from the
uncertainty in the $\Ds\to f_0(980)(\Kp\Km)\pip$ braching ratio.

Precise knowledge of the efficiency of reconstructing the tag is not
important, since it mostly cancels in the calculation of the partial
width ratio. However, the presence of two charged kaons in \dstophipi
events leads to an increased number of random tag candidates, compared
to \dstomunu events, which decreases the chances that the correct tag
is picked. The size of the correction for this effect to the
efficiency ratio ($\epsilon_{\phi\pi}/\epsilon_{\text{Sig}}$) is
determined to be \unit[$-1.4$]{\%} in simulated events.

To measure the effect of a difference between the \Dss momentum
spectrum in simulated and data events, \dsstodstophipi events are
selected in data with the \Dss momentum requirement removed. The
sample is purified by requiring the CM momentum of the charged pion to
be at least \unit[0.8]{\gevc}. The efficiency-corrected \Dss momentum
distribution in data is compared to that of \Dss in simulated
\dsstodstophipi events. A harder momentum spectrum is observed in
data.  The detection efficiencies for signal and \dsstodstophipi
events are re-evaluated after weighting simulated events to match the
\Dss momentum distribution measured in data. The correction to the
efficiency ratio is \unit[$+1.5$]{\%}.
 
With both corrections applied, the partial width ratio is determined
to be $\Gamma_{\mu\nu}/\Gamma_{\phi\pi} = (N/\epsilon)_{\text{Sig}}/(N/\epsilon)_{\phi\pi}\times\BR(\phi\to K^+K^-) = \PWRVal \pm
\PWRValStat(stat)$, with $\BR(\phi\to K^+K^-)=\unit[49.1]{\%}$ \cite{Eidelman:2004wy}.

A detailed discussion of the systematics can be found in~\cite{babardsmunu}. Using the \babar average for the branching ratio
$\BR(\dstophipi)=(4.71\pm0.46)\,\%\,$\cite{Aubert:2005xu}\cite{Aubert:2006nm},
the branching fraction $\BR(\dstomunu) = (\BRbabarVal \pm
\BRbabarValStat \pm \BRbabarValSyst \pm \BRbabarValNorm)\times10^{-3}$
and the decay constant $\fds = \unit[(\FDsbabarVal \pm
\FDsbabarValStat \pm \FDsbabarValSyst \pm \FDsbabarValNorm)]{\mev}$ are obtained.
The first and second errors are statistical and systematic,
respectively; the third is the uncertainty from $\BR(\dstophipi)$.

\subsection{$D_s \to \mu \nu_{\mu}$ at Belle}

The Belle analysis uses data corresponding to $548$ fb$^{-1}$ to study the decay  $D_s^+ \to \mu^+\nu_\mu$, using the full-reconstruction recoil method first established in the study of semileptonic $D$ mesons described above.

This analysis uses fully reconstructed events of the type $e^+e^-\rightarrow D^*_sD^{\pm,0}K^{\pm,0}X$, where $X$ can be any number of additional pions from fragmentation, and up to one photon\footnote{It has been found that events with additional kaons or more than one photon have a poor signal/background ratio and have been therefore excluded. }. The {\it tag side} consists of a $D$- and a $K$ meson (in any charge combination) while the {\it signal side} is a $D^*_s$ meson decaying to $D_s\gamma$. Reconstructing the tag side, and allowing any possible set of particles in $X$, the signal side is reconstructed in the recoil, using the known beam momentum.

\begin{figure*}[t]
  \begin{center}
    \includegraphics[width=6.75cm]{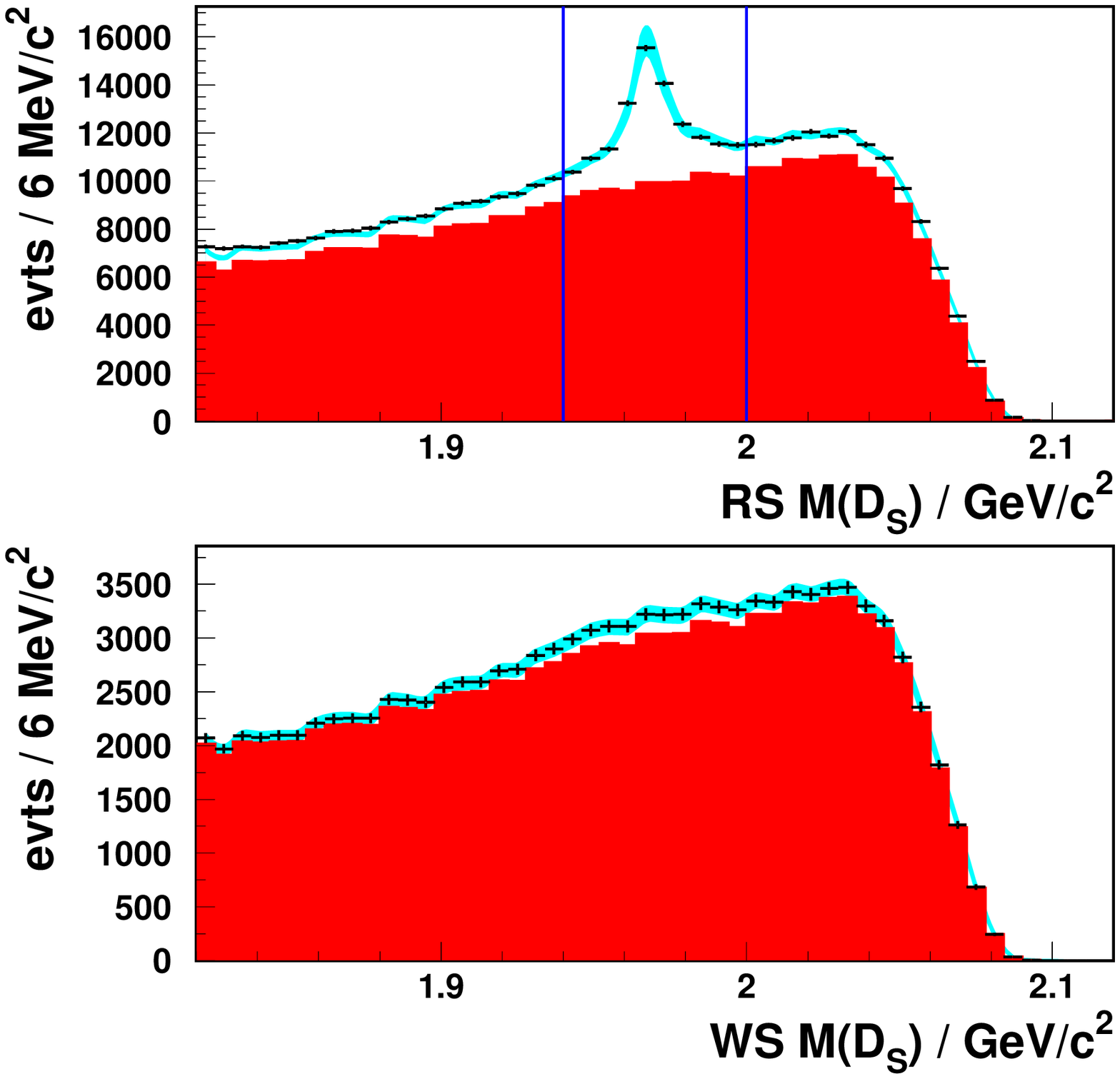}
    \includegraphics[width=6.75cm]{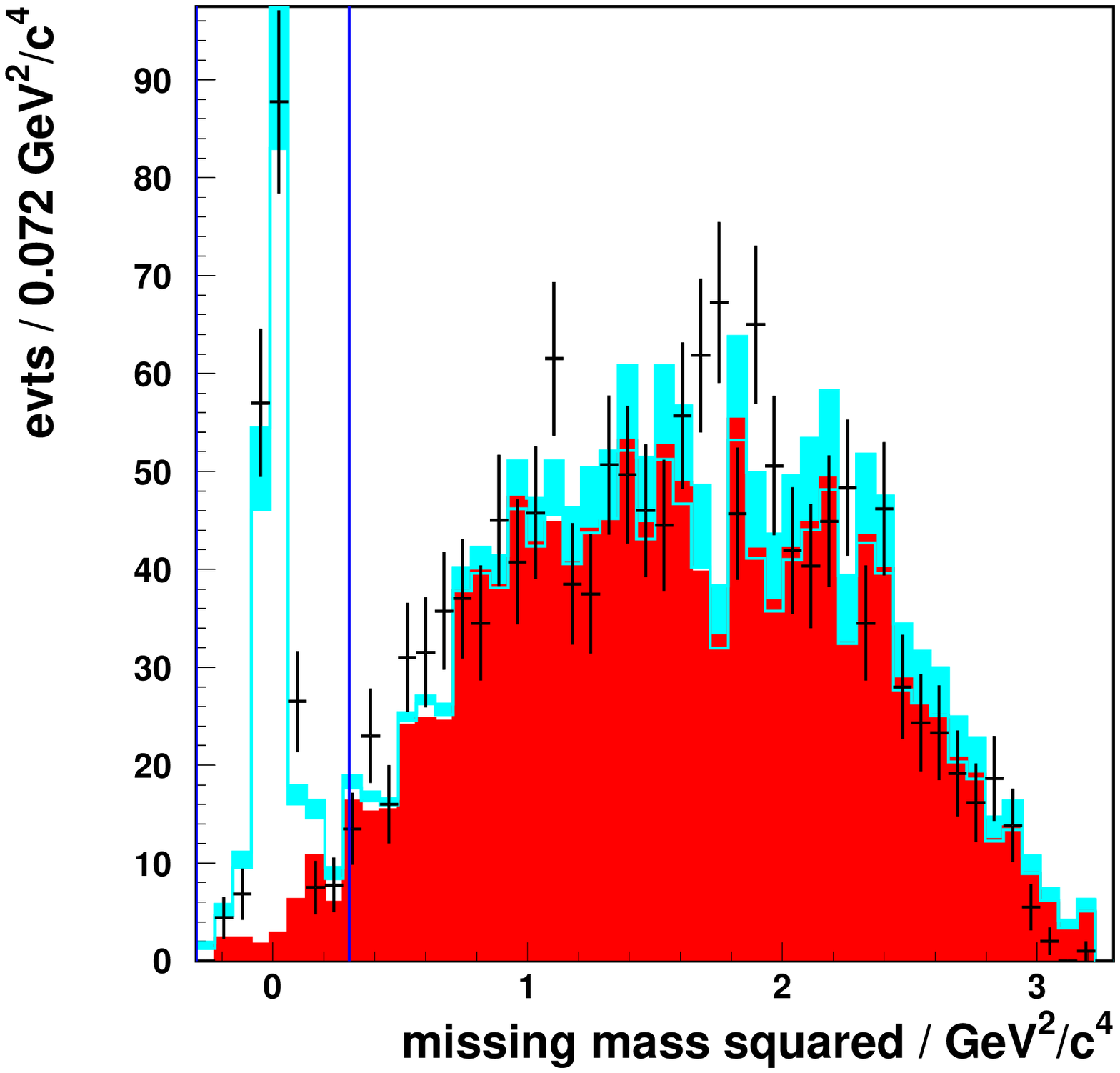}
  \end{center}
  \caption{Belle: Invariant mass spectrum of $D_s$-tags (left plot), and missing mass spectrum squared of $D_s \to \mu\nu$ candidates (right plot) in data with the selection criteria described in the text (points with error bars show statistical errors only). The red shaded areas show the fitted background, the cyan shaded bands show the fit with systematic uncertainties. The vertical lines indicate the signal regions.} 
\label{fig:m2dp0}
\end{figure*}

Tracks are detected with the CDC and the SVD. They are required to have  at least one associated hit in the SVD and an impact parameter with respect
 to the interaction point in the radial
 direction of less than $2$~cm and in the
 beam direction of less than $4$~cm.
 Tracks are also required to have momenta in the laboratory frame greater than $92$~MeV/$c$.
 A likelihood ratio for a given track to be a kaon or pion, which is required to be larger than $50\%$,
is obtained by utilising specific ionisation
 energy loss measurements made with the CDC, light yield
 measurements from the ACC, and time-of-flight information from the TOF.
 Lepton candidates are required to have momentum in the lab frame larger than 500
 MeV/$c$. For electron identification we use position, cluster energy,
 shower shape in the ECL, combined with track momentum and $dE/dx$ measurements in
 the CDC and hits in the ACC. For muon identification, we extrapolate
 the CDC track to the KLM and compare the measured range and
 transverse deviation in the KLM with the expected values.
Neutral pion candidates are reconstructed using photon pairs with invariant mass within $\pm 10$~MeV/$c^2$ of the nominal $\pi^0$ mass. Neutral kaon candidates are reconstructed using charged pion pairs within $\pm 30$~MeV/$c^2$ of the nominal $K^0$ mass.

Tag-side $D$ mesons (both charged and neutral) are reconstructed in $D \rightarrow K n \pi$ with $n=1,2,3$ 
(total branching fraction $\approx 25\%$). Mass windows have been optimised for each channel separately, and a combined mass- and vertex fit (requiring a confidence level greater than $0.1\%$) is applied to the $D$ meson to improve the momentum resolution.
$D^*_s$-candidates are not directly reconstructed, but searched for in the recoil of $DKX$, using the known beam momentum, by applying a mass window cut of $\pm 150$ MeV$/c^2$ around the nominal $D^*_s$ mass~\cite{ref:pdg06}. Since at this point in the reconstruction $X$ can be any set of remaining pions and photons, there are usually a large number of combinatorial possibilities. This number is reduced by requiring the presence of a photon that is consistent with the decay $D^*_s \to D_s\gamma$ where the $D_s$ has its nominal mass within a mass window of $\pm 150$ MeV$/c^2$. Further selection criteria are applied on the momentum of the primary $K$ meson in the $e^+e^-$ rest frame, $p_K$, which should be smaller than $2$~GeV/$c$; the momentum of the $D$ meson in the $e^+e^-$ rest frame, $p_D$, should be larger than $2$~GeV/$c$; the momentum of the $D^*_s$ meson in the $e^+e^-$ rest frame, $p_{D^*_s}$, is required to be larger than $3$~GeV/$c$ and the energy of the photon in $D^*_s \to D_s \gamma$, $E_{\gamma,D_s}$ in the lab frame, is required to be larger than $150$ MeV/$c^2$, irrespective of $\theta_\gamma$.
To further improve the recoil momentum resolution, inverse~\footnote{The fit is called {\it inverse} since it uses information from the mother and sister particles, rather than information about daughter particles as is usually the case.} mass-constrained vertex fits are then performed for the $D^*_s$ and $D_s$, requiring a confidence level greater than $1\%$. After all these selections are applied, the average number of combinatorial reconstruction possibilities is $\approx 2$ per event.
The sample is further divided into a right- (RS) and wrong-sign (WS) part, depending on the relative charges of the primary $K$ meson, the $D$ meson and that of the $K$ meson the $D$ meson decays into ($K_D$), compared to the charge of the $D^*_s$ meson, which is fixed by the total charge of the $X$ assuming overall charge conservation for the event.

Within this sample of $D_s$-tags, decays of the type $D_s \to \mu\nu_\mu$ are selected by requiring another charged track that is identified as a muon and has the same charge as the $D_s$ candidate. No additional charged particles are allowed in the event, and additional energy corresponding to neutral particles is required to be smaller than $1/n$~GeV where $n$ is the number of additional neutral particles. After these selections, in almost all cases only one combinatorial reconstruction possibility remains.
Figure \ref{fig:m2dp0} shows the mass spectra for $D_s$-tags and neutrino candidates.

$n_X$ is defined as the number of {\it primary} particles in the event, where {\it primary} means that the particle is not a daughter of any particle reconstructed in the event. The minimal value for $n_X$ is three corresponding to a $e^+e^- \to D^*_sDK$ event without any further particles from fragmentation. The upper limit for $n_X$ is determined by the reconstruction efficiency; Monte-Carlo (MC) shows that the number of reconstructed signal events is negligible for $n_X > 10$. 
As the efficiency very sensitively depends on $n_X$, it is crucial to use MC that correctly reflects the $n_X$ distribution observed in data. Unfortunately, the details of fragmentation processes are not very well understood, and standard MC~\cite{ref:bellegen} shows notable differences compared to data.
Furthermore, the true (generated) $n^T_X$ value differs from the reconstructed $n^R_X$, as particles can be lost or wrongly assigned. Thus the measured (reconstructed) $n^R_X$ distribution has to be deconvoluted so that the analysis can be done in bins of $n^T_X$ to avoid bias in the results.

To extract the number of $D_s$-tags as function of $n^T_X$ in data from the background, 2-dimensional histograms in $n^R_X$ (ranging from $3$ to $8$) and the invariant recoil mass $m_{D_s}$ are used. The signal shapes for different values of $n^T_X$ (ranging from $3$ to $9$~\footnote{The upper limit of $9$ is chosen because the bin $n^T_X = 9$ has some overlap with $n^R_X \le 8$.}) of the signal are modelled with generic MC (GMC)~\cite{ref:bellemc}, which has been filtered on the generator-level for events of the type $e^+e^- \to D^*_sDKX$. The weights of these components, $w_i^{D_s}, i=1..6$, are free parameters in the fit to data.
As a model for the background in RS, the WS data sample is used. Each slice of $n^R_X$ is fitted separately, adding another 6 free parameters. Since the WS-sample contains some signal ($\approx 10\%$ of the RS signal), these signal components (in slices of $n^T_X$) are also included in the fit as independent parameters (yielding a {\it negative} weight to compensate for the WS signal present in the data shapes). The fit is performed simultaneously with all these free parameters. As a crosscheck, the fit has also been performed using generic MC RS-sample backgrounds, which gives a negligible change in the results. A further crosscheck involved dividing the MC sample randomly into two halves, using the shapes of the first half to fit the signal in the second. The result as function of $n^T_X$, normalized to the amount of signal in the first half, fits to a flat line as $0.990 \pm 0.046$, which agrees well with the expectation of $1$.
The total number of reconstructed $D_s$-tags in data is calculated
as 
\begin{eqnarray}
&&N_{D_s}^{\rm rec}= \sum_{i=1}^6 w_i^{D_s}N_{D_s}^{{\rm GMC},i} ~,
\label{eq2}
\end{eqnarray}
where $N_{D_s}^{{\rm GMC},i}$ represents the total number of
reconstructed filtered GMC events that were generated 
in the $i$-th bin of $n^T_X$ (regardless of the reconstructed $n^R_X$) and $w^{D_s}_i$ is the fitted weight of this component. 

To fit the number of $D_s \to \mu\nu$-events as function of $n^T_X$, 2-dimensional histograms in $n^R_X$ and the missing mass squared $m^2_\nu$ are used. The shape of the signal is modelled with signal MC.
As MC studies show, the background under the $\mu\nu$-signal peak consists primarily of non-$D_s$ decays, semileptonic $D_s$ decays (where the additional hadrons have low momenta and remain undetected) and leptonic $\tau$ decays (where the $\tau$ decays to a muon and two neutrinos). Hadronic $D_s$ decays (with one hadron misidentfied as muon) are a rather small background component. 
Except for hadronic decays, which are negligible, all backgrounds are common to the $e\nu$-mode, which is suppressed by a factor of $O(10^{-5})$. Thus, the $e\nu$-sample provides a good model of the $\mu\nu$ background that has to be corrected only for kinematical and efficiency differences. Including this corrected shape in the fit,
the total number of fitted $\mu\nu$-events in data is given by
\begin{eqnarray}
&&N_{\mu\nu}^{\rm rec}= \sum_{i=1}^6 w_i^{\rm \mu\nu}N_{\mu\nu}^{{\rm SMC},i} ~,
\label{eqn3}
\end{eqnarray}
where $N_{\mu\nu}^{{\rm SMC},i}$ represents the total number of
reconstructed signal MC events that were generated 
in the $i$-th bin of $n^T_X$ (regardless of the reconstructed $n^R_X$) and $ w_i^{\rm \mu\nu}$ is the fitted weight of this component.

The numerical result for $N_{D_s}^{\rm rec}$ is $32100\pm870 (\mbox{stat})\pm1210 (\mbox{syst})$, that for $N_{\mu\nu}^{\rm rec}$ is $169\pm16 (\mbox{stat})\pm8 (\mbox{syst})$. The statistical uncertainties are due to statistics in the data signal, the systematic uncertainties due to statistics of the data background samples and those of the MC samples used. These errors include the non-negligble correlations between the $n^T_X$ bins.

As the branching fraction of $D_s \to \mu\nu$ used for the generation of generic MC is known, the branching fraction in data can be determined using the following formula:
\begin{equation}
  {\mathcal B}(D_s \to \mu\nu) = \frac{N_{\mu\nu}^{\rm rec}}{N_{\mu\nu}^{\rm GMC exp}} \cdot {\mathcal B}_{\rm generated}(D_s \to \mu\nu) ~,
\end{equation}
where ${\mathcal B}_{\rm generated}(D_s \to \mu\nu) = 0.0051$ and $N_{\mu\nu}^{\rm GMC exp}$ is the number of reconstructed $\mu\nu$-events in the generic MC, weighted according to the fit to data, i.e.
\begin{eqnarray}
&&N_{\mu\nu}^{\rm GMC exp}= \sum_{i=1}^6 w_i^{D_s}N_{\mu\nu}^{{\rm GMC},i} ~.
\end{eqnarray}
where $N_{\mu\nu}^{{\rm GMC},i}$ represents the total number of
reconstructed $\mu\nu$-events filtered from GMC that were generated 
in the $i$-th bin of $n^T_X$ (regardless of the reconstructed $n^R_X$). 

\begin{figure}
  \begin{center}
    \includegraphics[width=8.0cm]{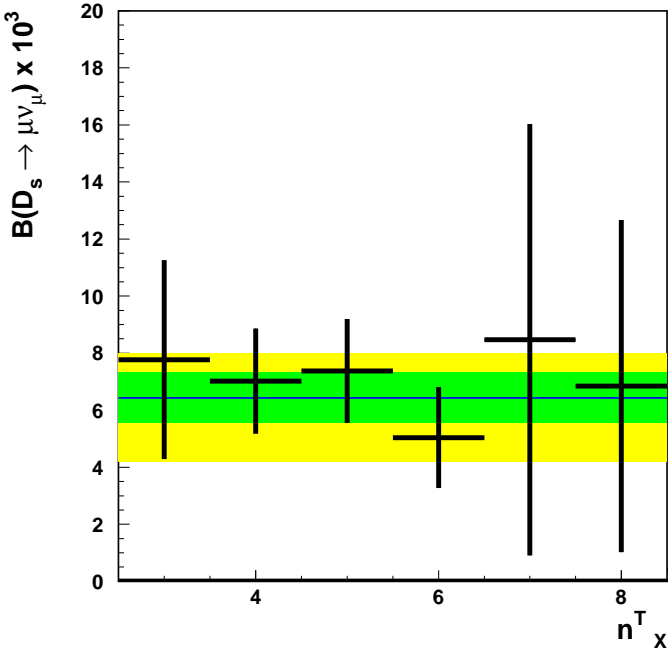}
  \end{center}
  \caption{Belle: ${\mathcal B}(D_s \to \mu\nu)$ as a function of $n^T_X$; final result is shown as the green shaded region. For comparison, the PDG value and its error is shown as yellow shaded region in the background.}
\label{fig:getbr}
\end{figure}

Figure~\ref{fig:getbr} shows the branching fraction determined in bins of $n_X^T$ (using correlated fit results). The result is stable within errors in $n_X$; note that the errors shown for the $n_X$ bins are the total errors, including correlation. The final result is:
\begin{equation}
  {\mathcal B}(D_s \to \mu\nu) = (6.44 \pm 0.76 (\mbox{stat}) \pm 0.52 (\mbox{syst})) \cdot 10^{-3}
\end{equation}
The statistical error reflects the statistics of the signal sample. The systematic error is dominated by statistical uncertainties due to background samples from data and MC samples ($0.29$) and the statistical uncertainty on $N_{\mu\nu}^{\rm GMC exp}$ ($0.41$). Since the branching fraction is determined by calculating a ratio of the signal yield to the number of $D_s$-tags, systematics in the reconstruction of the tag side cancel; the only remaining systematics are due to the tracking and identification of the muon, which have been estimated as $2\%$, contributing $0.13$ to the total systematic error. 
As a crosscheck, also the branching fraction for $n^T_X \le 6$ has been determined as $(6.54 \pm 0.76 (\mbox{stat}) \pm 0.54 (\mbox{syst})) \cdot 10^{-3}$, which agrees nicely with the result given above including all available $n^T_X$ bins. 

The decay constant $f_{D_s}$, using Eqn.~\ref{eqn:gammunu} and recent values from PDG~\cite{ref:pdg06} yields
\begin{equation}
	f_{D_s} = 275 \pm 16 (\mbox{stat}) \pm 12 (\mbox{syst}) \mbox{MeV}. 
\end{equation}
.

\section{Summary: Overview and Comparison}

\begin{table*}[t]
\caption{Overview of results obtained by BaBar and Belle experiments, compared to theoretical expectations (where available); errors are statistical (first) and systematic (second).}
{\begin{tabular}{lc|r|r|r|r} \toprule
decay mode & parameter & prediction & BaBar result & Belle result & difference \\
\hline
$D \to Ke\nu_e$ & BF (\%) &  & $3.522 \pm 0.027 \pm 0.079$ & $3.45 \pm 0.10 \pm 0.19$ & $0.3\sigma$ \\
 & z-expansion, $a_0$ & & $2.98 \pm 0.01 \pm 0.04$ & n/a \\
 & z-expansion, $a_1$ & & $-2.5 \pm 0.2 \pm 0.2$ & n/a \\
 & z-expansion, $a_2$ & & $0.6 \pm 6.0 \pm 5.0$ & n/a \\
 & $m_{\rm{pole}}$ (GeV/$c^2$) & $2.112 (=m_{D^*_s})$ & $1.884 \pm 0.012 \pm 0.015$ & $1.82 \pm 0.04 \pm 0.03$ & $1.2\sigma$ \\
 & $\alpha$ & $0.50\pm0.04$ & $0.377 \pm 0.023 \pm 0.029$ & $0.52 \pm 0.08 \pm 0.06$  & $1.3\sigma$\\
 & $f_+(0)$ & $0.73(3)(7)$ & $0.727 \pm 0.007 \pm 0.009$ & $0.695 \pm 0.007 \pm 0.022$  & $1.3\sigma$\\
$D \to K\mu\nu_\mu$ & BF (\%) &  & n/a & $3.45 \pm 0.10 \pm 0.21$ \\
 & $m_{\rm{pole}}$ (GeV/$c^2$) & $2.112 (=m_{D^*_s})$ & n/a & {\small included in $Ke\nu_e$} \\
 & $\alpha$ & $0.50\pm0.04$ & n/a & {\small results shown} \\
 & $f_+(0)$ & $0.73(3)(7)$ & n/a & {\small above} \\
$D \to \pi e\nu_e$ & BF (\%) &  & n/a & $0.279 \pm 0.027 \pm 0.016$ \\
 & $m_{\rm{pole}}$ (GeV/$c^2$) & $2.010 (=m_{D^*})$ & n/a & $1.97 \pm 0.08 \pm 0.04$ \\
 & $\alpha$ & $0.44\pm0.04$ & n/a & $0.10 \pm 0.21 \pm 0.10$ \\
 & $f_+(0)$ & $0.64(3)(6)$ & n/a & $0.624 \pm 0.020 \pm 0.030$ \\
$D \to \pi\mu\nu_\mu$ & BF (\%) &  & n/a & $0.231 \pm 0.026 \pm 0.019$ \\
 & $m_{\rm{pole}}$ (GeV/$c^2$) & $2.010 (=m_{D^*})$ & n/a & {\small included in $\pi e\nu_e$} \\
 & $\alpha$ & $0.44\pm0.04$ & n/a & {\small results shown} \\
 & $f_+(0)$ & $0.64(3)(6)$ & n/a & {\small above} \\

$D_s \to \phi e\nu_e$ & $r_2$ ($m_{A,V}$ fixed) & & $0.705 \pm 0.056 \pm 0.029$ & n/a \\
 & $r_V$ ($m_{A,V}$ fixed) & & $1.636 \pm 0.067 \pm 0.038$ & n/a \\
 & $r_2$ ($m_V$ fixed) & & $0.711 \pm 0.111 \pm 0.096$ & n/a \\
 & $r_V$ ($m_V$ fixed) & & $1.633 \pm 0.081 \pm 0.068$ & n/a \\
 & $m_A$ (GeV/$c^2$) ($m_V$ fixed) & & $2.53 ^{+0.54}_{-0.35} \pm 0.54$ & n/a \\
$D_s \to \mu\nu_\mu$ & BF ($10^{-3}$) & & $6.74 \pm 0.83 \pm 0.71$ & $6.44 \pm 0.76 \pm 0.52$ & $0.2\sigma$ \\
 & $f_{D_s}$ (MeV) & $249 \pm 3 \pm 16$ & $283 \pm 17 \pm 16$ & $275 \pm 16 \pm 12$  & $0.3\sigma$\\
\hline
\end{tabular}
\label{tab:overview}} \end{table*}

Studies in the charm sector are notoriously difficult for experiments running at much higher than threshold energy; still both Belle and \babar present an interesting variety of results on (semi)leptonic charm decays.
While Belle concentrated on fully reconstructed (and consequently tagged) events, \babar preferred methods with partially reconstructed events, and uses tag information only for its $D_s \to \mu\nu_\mu$ analysis. 

The advantage of Belle's approach is a very effective background suppression and an excellent neutrino momentum resolution which can compete with results achieved at experiments operating  at threshold energy like Cleo-c~\cite{cleoc}. It also allows absolute measurements, by normalization to the number of $D_{(s)}$ tags.

However, \babar's approach is significantly more efficient in terms of event statistics. In the case of the $D^0 \to Ke\nu_e$ analysis, despite higher backgrounds, this results in a much better accuracy of measurements. For the $D_s \to \mu\nu_\mu$ analysis, the advantage of higher statistics is more or less equalled by the disadvantage of larger backgrounds, which eventually gives somewhat larger errors than Belle's.

In any case, it is a valuable crosscheck to have rather different experimental approaches at different experiments.
Table \ref{tab:overview} summarizes all results by Belle and \babar.

Within the semileptonic channels discussed in this review, only the mode $D^0 \to Ke\nu_e$ has been studied by both \babar and Belle, and can be compared. 
The measured branching fractions agree well within errors, larger differences are seen in the form factor measurements. However, these differences do not exceed $1.3\sigma$, and could be due to the more complicated systematics of the fits involved. The results are much more precise than those of previous experiments, and also well compatible with other recent results~\cite{cleo3,focus}.
Obviously, the untagged, partial reconstruction of \babar has clearly smaller erros, even though it suffers a large uncertainty due to the normalizing channel used, which is the dominating part of its systematic error. Belle does an absolute measurement, but also is clearly limited by systematics. Thus in neither case a significant further improvement of the measurements can be expected with more data accumulated, without also further developing the experimental methods. 

The other mode where results can be compared is $D_s \to \mu\nu_\mu$. The results agree very well with each other. Here Belle profits from its full reconstruction method, and has somewhat, but not dramatically smaller errors. Both results are well compatible with other recent results, and still compatible with theoretical predictions, which tend to be lower by $\approx 2.5\sigma_{\rm{exp}}$, but bear some uncertainties as well~\cite{theofds}.
In both experiments, statistical and systematic error are of the same order. Considering the fact that part of the systematic error is due to the size of control samples which will get larger with more statistics, there is some room for further improvements once the full data sets of the experiments are available.

\begin{acknowledgments}
We wish to thank Arantza Oyanguren, Justine Serrano and Paul Jackson from \babar for providing material to compile this talk. Text excerpts from \babar publications covering the described analyses have been used to compile this review.
As far as my own collaboration Belle is concerned, the thanks go to
the KEKB group for the excellent operation of the
accelerator, the KEK cryogenics group for the efficient
operation of the solenoid, and the KEK computer group and
the National Institute of Informatics for valuable computing
and Super-SINET network support. We acknowledge support from
the Ministry of Education, Culture, Sports, Science, and
Technology of Japan and the Japan Society for the Promotion
of Science; the Australian Research Council and the
Australian Department of Education, Science and Training;
the National Science Foundation of China under
contract No.~10575109 and 10775142; the Department of
Science and Technology of India; 
the BK21 program of the Ministry of Education of Korea, 
the CHEP SRC program and Basic Research program 
(grant No.~R01-2005-000-10089-0) of the Korea Science and
Engineering Foundation, and the Pure Basic Research Group 
program of the Korea Research Foundation; 
the Polish State Committee for Scientific Research; 
the Ministry of Education and Science of the Russian
Federation and the Russian Federal Agency for Atomic Energy;
the Slovenian Research Agency;  the Swiss
National Science Foundation; the National Science Council
and the Ministry of Education of Taiwan; and the U.S.\
Department of Energy.

\end{acknowledgments}

\bigskip 


\begin{thebibliography}{99} 

\bibitem{ref:babar}
B.\ Aubert {\em et al.} [\babar Collaboration], 
Phys.\ Rev.\ {\bf D66}, 032003 (2002).
\bibitem{Belle} A.\ Abashian {\it et al.} [Belle Collaboration], Nucl. Instr. and Meth. A {\bf 479}, 117 (2002).
\bibitem{Kronfeld:2006sk}
  A.~S.~Kronfeld [Fermilab Lattice Collaboration],
  J.\ Phys.\ Conf.\ Ser.\  {\bf 46} (2006) 147
  [arXiv:hep-lat/0607011].
\bibitem{svd2} Z.~Natkaniec {\it et al.} [Belle Collaboration], Nucl. Instr. and Meth. A {\bf 560}, 1 (2006).
\bibitem{ref:unquenched} M.\ Okamoto {\it et al.,}  Nucl.\ Phys.\ Proc.\ Suppl. \ {\bf 129}, 334 (2004). 
\bibitem{ref:unquenched2} C.\ Aubin {\it et al.,} [Fermilab Lattice Collaboration, MILC Collaboration and HPQCD Collaboration], Phys.\ Rev.\ Lett. \ {\bf 94}, 011601 (2005). 
\bibitem{ref:quenched}
  A.~Abada, D.~Becirevic, P.~Boucaud, J.~P.~Leroy, V.~Lubicz and F.~Mescia,
  Nucl.\ Phys.\ B {\bf 619}, 565 (2001).
\bibitem{ref:2a} G.\ Amoros, S.\ Noguera, J.\ Portoles, Eur.\ Phys.\ J.\ {\bf C27}, 243 (2003). 
\bibitem{ref:bk}
D. Becirevic and A.B. Kaidalov, Phys.~Lett.~B~{\bf 478},~417~(2000) [arXiv:hep-ph/9904490].
\bibitem{ref:hill1}
R. J. Hill, Proceedings of 4th Flavor Physics and CP Violation Conference (FPCP 2006), Vancouver, British Columbia, Canada, 9-12 Apr 2006, pp 027,
[arXiv:hep-ph/0606023].
\bibitem{ref:belle}
L. Widhalm  {\it et al.} [Belle collaboration], Phys. Rev. Lett. {\bf 97}, 061804 (2006).
\bibitem{ref:bellegen} see http://www.lns.cornell.edu/public/CLEO/soft/QQ.
\bibitem{ref:bellemc} R.~Brun {\it et al.,} GEANT 3.21, CERN Report DD/EE/84-1, (1984).
\bibitem{babarkenu} 
  B.~Aubert {\it et al.}  [\babar Collaboration],
  arXiv:0704.0020 [hep-ex].
\bibitem{ref:svd}
A. H\"{o}cker and V. Kartvelishvili, Nucl. Instrum. Methods {\bf A372}, 469 (1996).
\bibitem{ref:pdg06}
W.-M. Yao {\em et al.}, Review of Particle Physics,  Journal of Physics G {\bf 33}, 1 (2006).
\bibitem{ref:KS}
   G.Kopp, G. Kramer, G.A.~Schuler and W.F. Palmer, Z.~Phys.~C~{\bf 48},~327~(1990).
\bibitem{babarphienu} 
  B.~Aubert {\it et al.}  [\babar Collaboration],
  [arXiv:hep-ex/0607085].
\bibitem{ref:ks2}
J.G. Koerner and G.A. Schuler, Z.~Phys.~C~{\bf 38},~511~(1988);
Erratum-ibid~C~{\bf 41},~690~(1989).\\
M. Bauer and M. Wirbel, Z.~Phys.~C~{\bf 42},~671~(1989).
\bibitem{ref:fk} S.~Fajfer, J.~Kamenik, \newblock Phys.\ Rev.\ D {\bf 72}, 034029 (2005).

\bibitem{ref:wiss}
J. Wiss, ``Recent results on fully leptonic and semileptonic charm decays'',
FPCP Conference, Vancouver 2006 [arXiv:hep-ex/0605030].
\bibitem{babardsmunu}
  B.~Aubert {\it et al.}  [\babar Collaboration],
  Phys.\ Rev.\ Lett.\  {\bf 98} (2007) 141801
  [arXiv:hep-ex/0607094].
\bibitem{Chadha:1998zh} M.~Chadha {\em et~al.} [CLEO Collaboration],
  \newblock Phys.\ Rev.\ D {\bf 58}, 032002 (1998).
\bibitem{Eidelman:2004wy} S.~Eidelman {\em
    et~al.} [Particle Data Group], \newblock Phys.\ Lett.\ B {\bf 592}, 1 (2004).
\bibitem{Aubert:2005xu} B.~Aubert {\em et~al.} [\babar Collaboration],
  \newblock Phys.\ Rev.\ D {\bf 71}, 091104 (2005).
\bibitem{Aubert:2006nm} B.~Aubert {\em
    et~al.} [\babar Collaboration], \newblock hep-ex/0605036,\ to\ appear\ in\ Phys.\ Rev.\ 
  D\ -\ Rapid\ Communications.
\bibitem{cleoc} T.\ Pedlar {\it et al.} (CLEO-c Collab.), hep-ex/0704.0437
\bibitem{cleo3}
G. S. Huang {\it et al.} [CLEO collaboration], Phys. Rev. Lett. {\bf 94}, 011802 (2005).
\bibitem{focus}
J. M. Link {\it et al.} [FOCUS collaboration], Phys. Lett. {\bf B607}, 233 (2005).
\bibitem{theofds} C.~Aubin {\em et~al.}, \newblock Phys.\ Rev.\
  Lett. {\bf 95}, 122002 (2005).
\end{thebibliography}
\end{document}